\documentclass{svmult}

\usepackage{makeidx}
\usepackage{imakeidx}
\usepackage{graphicx}
\usepackage[colorlinks=true,linkcolor=blue,citecolor=blue,urlcolor=blue]{hyperref}
\usepackage{amsmath}
\usepackage{amssymb}
\usepackage{siunitx}

\newcommand{\citeasnoun}[1]{Ref.~\cite{#1}}

\newcommand{\figref}[1]{Fig.~\ref{fig:#1}}
\renewcommand{\eqref}[1]{Eq.~(\ref{eq:#1})}
\newcommand{\Eqref}[1]{Equation~(\ref{eq:#1})}
\newcommand{\eqreftwo}[2]{Eqs.~(\ref{eq:#1},\ref{eq:#2})}
\newcommand{\eqrefrange}[2]{Eqs.~(\ref{eq:#1})--(\ref{eq:#2})}
\newcommand{\secref}[1]{Sec.~\ref{sec:#1}}

\newcommand*{\GG}{\mathbb{G}}
\newcommand{\vect}[1]{\mathbf{#1}}
\newcommand*{\alphat}{\boldsymbol{\alpha}}
\newcommand*{\xv}{\vect{x}}
\newcommand*{\Ev}{\vect{E}}
\newcommand*{\ev}{\vect{e}}
\newcommand*{\Hv}{\vect{H}}
\newcommand*{\Einc}{\Ev_{\rm inc}}
\newcommand*{\Escat}{\Ev_{\rm scat}}
\newcommand*{\einc}{\ev_{\rm inc}}

\newcommand*{\Pv}{\vect{P}}
\newcommand*{\pv}{\vect{p}}
\newcommand*{\Jv}{\vect{J}}
\newcommand*{\TT}{\mathbb{T}}
\newcommand*{\DD}{\mathbb{D}}
\newcommand*{\II}{\mathbb{I}}
\newcommand*{\Pext}{P_{\rm ext}}
\newcommand*{\Pscat}{P_{\rm scat}}
\newcommand*{\Pabs}{P_{\rm abs}}
\newcommand*{\dxv}{\,{\rm d}\xv}
\newcommand*{\dw}{\,{\rm d}\omega}

\newcommand*{\nhat}{\hat{\vect{n}}}
\newcommand*{\wt}{\tilde{\omega}}

\renewcommand{\Re}{\operatorname{Re}}
\renewcommand{\Im}{\operatorname{Im}}
\newcommand{\Tr}{\operatorname{Tr}}

\makeindex 

\begin{document}

\title*{Fundamental limits to near-field optical response}
\author{Owen D. Miller}
\institute{Owen D. Miller \at Dept. of Applied Physics, Yale University, \email{owen.miller@yale.edu}}

\maketitle
Excerpted from \textit{Advances in Near-Field Optics}, R. Gordon, ed., forthcoming

\abstract*{Near-field optics is an exciting frontier of photonics and plasmonics. The tandem of strongly localized fields and enhanced emission rates offers significant opportunities for wide-ranging applications, while also creating basic questions: How large can such enhancements be? To what extent do material losses inhibit optimal response? Over what bandwidths can these effects be sustained? This chapter surveys theoretical techniques for answering these questions. We start with physical intuition and mathematical definitions of the response functions of interest, after which we describe the general theoretical techniques for bounding such functions. Finally, we apply those techniques specifically to near-field optics, for which we describe known bounds, optimal designs, and open questions.}

\section{Introduction}
Near-field optics\index{near field} is an exciting frontier of photonics and plasmonics\index{plasmons}. The near field is the region of space within much less than one electromagnetic wavelength of a source, and ``near-field optics'' refers to the phenomena that arise when optical-frequency sources interact with material structures in their near field. Free-space waves exhibit neglible variations over such small length scales, which might lead one to think this regime simply reduces to classical electrostatics and circuit theory. A new twist in the optical near field is the emergence of \textbf{polaritons}\index{polaritons}, modes that arise near the interfaces between negative- and positive-permittivity materials~\cite{Maier2005}. Polaritons emerge from an interplay of geometry and material susceptibility, instead of geometry and wave interference, to confine optical waves. Freedom from wave-interference requirements leads to a striking possibility: resonant fields whose size (spatial confinement) is \emph{decoupled} from its wavelength. Highly confined polaritons enable two reciprocal effects: incoming free-space waves can be concentrated to spatial regions much smaller than the the electromagnetic wavelength (well below the diffraction limit), and, conversely, that patterned materials close to a dipolar emitter can significantly amplify outgoing radiation.

The tandem of strongly localized fields and enhanced emission rates offers significant opportunities for applications including spectroscopy~\cite{Betzig1993,Taminiau2008}, nanolasers~\cite{Khajavikhan2012}, coherent plasmon\index{plasmons} generation~\cite{Oulton2009}, and broadband single-photon sources~\cite{Maksymov2010}. It also generates fundamental questions: How large can such enhancements be? Are there limits to field localization? All known polaritonic materials have significant or at least non-trivial amounts of material loss; to what extent does the loss affect these quantities? Over what bandwidths can these effects be sustained?

This chapter surveys theoretical techniques for answering these questions. The same features that make the near field appealing also make it theoretically challenging: there are not fixed photon flows, modal descriptions require exquisite care, and analytical descriptions are not possible except in the simplest high-symmetry scenarios. Over the past decade, thankfully, there has been a surge of interest in identifying what is possible in these systems. One key to the success of these approaches is to not attempt to develop models that apply to every possible instance of a given scattering scenario, but instead to develop techniques that identify \emph{bounds} to the \emph{extreme possibilities} of each scattering scenario. In this chapter, we describe these techniques in detail. We start with physical intuition and mathematical definitions of the response functions of interest (\secref{ResponseFunctions}), after which we describe the general theoretical techniques for bounding such functions (\secref{BoundApproaches}). Finally, we apply those techniques specifically to near-field\index{near field} optics, for which we describe known bounds, optimal designs, and open questions (\secref{NearFieldLimits}).

\section{Near-field optical response functions}
\label{sec:ResponseFunctions}
\begin{figure}
    \includegraphics[width=1\textwidth]{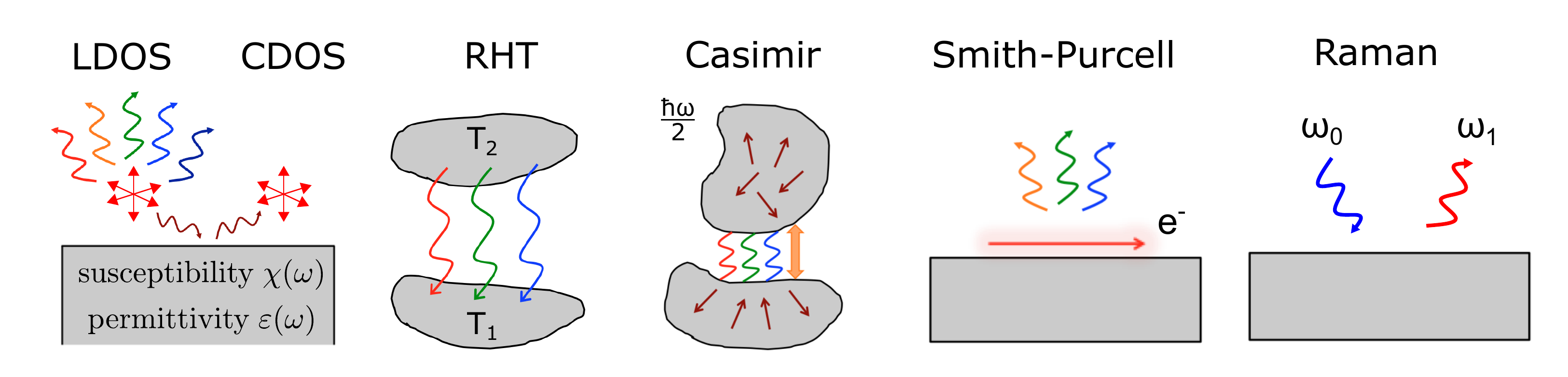}
    \centering
    \caption{An array of near-field\index{near field} optical response functions of broad interest. (Adapted from \citeasnoun{Shim2019}.)}
    \label{fig:rfx}
\end{figure}

In this section we summarize the background intuition and mathematical equations describing six key near-field\index{near field} optical response functions: local density of states (\secref{LDOS}), which is proportional to the radiation of a single dipolar current, free-electron radiation\index{free-electron radiation} (\secref{FER}), which is the collective radiation of a line of current created by an electron beam, the cross density of states (\secref{CDOS}), which measures modal or emission correlations across different spatial locations, surface-enhanced Raman scattering (\secref{SERS}), which is the simultaneous enhancement of incident radiation and outgoing luminescence, typically for imaging or sensing applications, near-field radiative heat transfer (\secref{NFRHT}), which is the transfer of radiative energy from a hot body to a cold one, at near-field separations, and mode volume\index{mode volume} (\secref{modev}), which refers to the spatial confinement of a resonant mode. Many of these response functions are depicted in \figref{rfx}.

\subsection{LDOS}
\label{sec:LDOS}
The first and arguably most important near-field\index{near field} response quantity is the \emph{local density of states} (LDOS)\index{local density of states (LDOS)}. The central role of LDOS is a result of the extent to which it underpins many connected ideas in near-field\index{near field} optics~\cite{Joulain2003}.

The first connection is to the power radiated by a dipole. In general, the work per time done by a field $\Ev$ on a current $\Jv$ in a volume $V$ is given by $(1/2) \Re\int_V \Jv^* \cdot \Ev$. This is a generalized version of Watt's Law in circuit theory, and it encodes the work done by the electric field mediating the electric force on the charges in the current, across a distance traveled by the charges given by the product of their speed and the time interval of interest. By Newton's second law, the work per time done by a current $\Jv$ on a field $\Ev$ is the negative of the expression above, $-(1/2) \Re\int_V \Jv^* \cdot \Ev$. We can convert the current density $\Jv$ to a dipole density $\Pv$ by the relation $\Jv = \partial \Pv / \partial t = -i\omega\Pv$ for harmonic frequency $\omega$ ($e^{-i\omega t}$ convention). Then the power radiated by a dipole at $\xv_0$ with dipole moment $\pv$ (and therefore dipole density $\Pv = \pv \delta(\xv - \xv_0)$) is
\begin{align}
    P_{\rm rad} &= -\frac{1}{2} \Re \int_V \Jv^* \cdot \Ev \,{\rm d}\xv \nonumber \\
                &= \frac{\omega}{2} \Im \int_V \Pv^* \cdot \Ev \,{\rm d}\xv \nonumber \\
                &= \frac{\omega}{2} \Im \left[\pv^* \cdot \Ev(\xv_0)\right].\nonumber
\end{align}
The electric field at $\xv_0$, $\Ev(\xv_0)$, is the field produced by a delta-function dipole source, which exactly coincides with the dyadic Green's function (GF) $\GG$, evaluated at $\xv_0$ from a source at $\xv_0$, multiplied by the dipole moment $\pv$, giving:
\begin{align}
    P_{\rm rad} &= \frac{\omega}{2} \Im \left[ \pv^* \cdot \GG\left(\xv_0,\xv_0\right) \pv \right]. \nonumber
\end{align}
The imaginary part of a complex number of the form $z^\dagger A z$ is $\Im(z^\dagger A z) = z^\dagger (\Im A) z$ by symmetry, where $\Im A$ refers to the anti-Hermitian part of $A$ ($\Im A = (A - A^{\dagger})/2i$). So we have
\begin{align}
    P_{\rm rad} = \frac{\omega}{2} \pv^\dagger \left[ \Im \GG\left(\xv_0,\xv_0\right)\right] \pv. \label{eq:PRadDip}
\end{align}
This result gives us the first key near-field\index{near field} response function, the imaginary part of the Green's function evaluated at the source position,
\begin{align}
    \Im \GG(\xv_0,\xv_0),
\end{align}
which is proportional to the radiation rate of an electric dipole into any environment.

{\bf Spontaneous emission}\index{spontaneous emission} typically occurs via electric-dipole transitions in atomic or molecular systems, so the rate of spontaneous emission is governed by the imaginary part of the GF. It has been recognized for many decades that this rate is not an immutable constant, but a function of the environment. Just as specifying the amplitude of a current or voltage source in a circuit does not dictate the power delivered by the source, which depends on the impedance of the load, specifying the amplitude of a dipole moment does not dictate the power it delivers to its electromagnetic environment. This fact inspired the concept of a photonic bandgap~\cite{Yablonovitch1987} and photonic crystals~\cite{Yablonovitch1994,Joannopoulos2011}, with the goal for \emph{inhibiting} spontaneous emission\index{spontaneous emission}, originally to avoid laser power loss. It has conversely inspired significant effort towards \emph{amplifying} spontaneous emission\index{spontaneous emission}, for applications such as single-molecule imaging~\cite{Betzig1993,Taminiau2008}. An early recognition of this fact came from Purcell, who noted that an emitter radiating into a single-photonic-mode environment would have an altered spontaneous emission rate~\cite{Purcell1946}. Purcell recognized that for a single-mode resonator with quality factor $Q$ and mode volume\index{mode volume} $V$, the density of states (per unit volume and per unit frequency) becomes $(Q/\omega)/V$. The relative change of the spontaneous-emission rate is the {\bf Purcell factor}\index{Purcell factor}, which is proportional to $\lambda^3 Q/V$.

Purcell derived this expression in the context of enhancing magnetic-dipole transitions in spin systems, but exactly the same argument applies to electric-dipole transitions, where it is most used today. This expression drives many modern investigations of high-quality-factor and/or small-mode-volume cavity design~\cite{Pelton2002,Kippenberg2004,Stoltz2005,Lipson05,Liang2013,Choi2017,Hu2018}, to reach the largest Purcell enhancement\index{Purcell factor} possible. It can be generalized to multi-mode, high-$Q$ systems: if each mode has mode field $\Ev_i$, center frequency $\omega_i$, and linewidth (half-width at half-maximum) $\gamma_i$, the power radiated by a dipole with moment $\pv$ located at position $\xv_0$ is~\cite{taflove_oskooi_johnson_2013}
\begin{align}
    P_{\rm rad} \approx \frac{\omega^2}{4} \sum_i \frac{\gamma_i |\Ev_i^\dagger(\xv_0) \pv|^2}{(\omega - \omega_i)^2 + \gamma_i^2}
    \label{eq:lorentz}
\end{align}
In the limit of infinite $Q$, the Lorentzian lineshapes become delta functions, and the summation simplifies to delta functions multiplied by the overlap of modal fields with the dipole moment. The overlap of each mode with the dipole is a measure of the relative modal energy concentration at that particular point in space. Hence the overall summation can be understood as a {\bf local density of states}\index{local density of states (LDOS)}, or LDOS (with appropriate prefactors). The power radiated by a dipole into an electromagnetic environment, then, is directly proportional to the local density of electromagnetic modes; inserting the correct prefactors leads to an LDOS expression in terms of $\Im\GG$~\cite{Joulain2003,Economou2006,Wijnands1997, xu_lee_yariv_2000}:
\begin{align}
    \textrm{LDOS}(\omega,\xv) = \frac{1}{\pi\omega} \Tr \Im \GG(\xv_0,\xv_0),
    \label{eq:LDOSe}
\end{align}
where the trace encodes a summation over all independent polarizations. (Note that e.g. \citeasnoun{Joulain2003} defines the Green's function with an extra $1/\omega^2$ factor, which leads to $\omega$ in the numerator of their analog to \eqref{LDOSe}.) In free space, the LDOS\index{local density of states (LDOS)} coincides with the density of states (as there are no spatial variations), and is given by $\textrm{LDOS}(\omega) = \omega^2 / 2\pi^2 c^3$. Technically, the expression of \eqref{LDOSe} is the \emph{electric} LDOS; one can similarly define a magnetic LDOS through a summation over the relative magnetic-field strengths, or more generally by the power radiated by a magnetic dipole. For a magnetic Green's function $\GG^{(HM)}$, denoting the magnetic field from a magnetic-dipole source, the magnetic LDOS is~\cite{Joulain2003}
\begin{align}
    \textrm{LDOS}^{(m)}(\omega,\xv) = \frac{1}{\pi\omega} \Tr \Im \GG^{(HM)}(\xv_0,\xv_0).
        \label{eq:LDOSm}
\end{align}
The sum of \eqref{LDOSe} and \eqref{LDOSm} is referred to as the \emph{total} LDOS, representing the totality of electric- and magnetic-field energy localized to a point $\xv_0$, at frequency $\omega$, over all modes. (Significant alterations to the modal-decomposition expressions are needed, for example, in plasmonic\index{plasmons} (and polaritonic\index{polaritons}) systems~\cite{Sauvan2013,Lalanne2018}.) Such descriptions are mathematically accurate only in the high-quality-factor limit, but the dipole-radiation interpretation generalizes to any linear scattering scenario.

To summarize, the imaginary part of the Green's function, $\Im \GG(\xv_0,\xv_0)$, is a measure of the power radiated by elecric and/or magnetic dipoles in an arbitrary environment, which is proportional to the spontaneous-emission\index{spontaneous emission} rate of a dipolar emitter, and it encapsulates the Purcell factor\index{Purcell factor}, particularly the ratio $Q/V$, of high-quality-factor modes that concentrate energy at that point. We have extensively described LDOS due to its versatility and cross-cutting nature. The following quantities have more focused and niche applications, and can be described more concisely.

\subsection{Free-electron radiation}
\label{sec:FER}
Radiation by a free-electron beam\index{free-electron radiation} is closely related to LDOS, with the key distinction being that the current distribution is now a line source. An electron (charge $-e$) propagating through free space at constant velocity $v\hat{\xv}$ comprises a free current density $\Jv(\vect{r},t) = -\hat{\xv} ev \delta(y) \delta(z) \delta(x-vt)$, which generates a frequency-dependent incident field~\cite{GarciaDeAbajo2010}
\begin{align}
    \Einc = \frac{e\kappa_p e^{ik_v x}}{2\pi\omega\varepsilon_0} \left[\hat{\xv} i \kappa_{\rho} K_0 (\kappa_{\rho} \rho) - \hat{\boldsymbol{\rho}} k_v K_1(\kappa_{\rho}\rho)\right],
    \label{eq:EincFE}
\end{align}
written in cylindrical coordinates $(x,\rho,\theta)$, where $K_n$ is the modified Bessel function of the second kind, $k_v = \omega/v$, and $\kappa_{\rho} = \sqrt{k_v^2 - k^2} = k/\beta\gamma$ ($k=\omega/c$, free-space wavevector; $\gamma = 1/\sqrt{1-\beta^2}$, Lorentz factor). Then photon emission and energy loss of free electrons interacting with nearby scatterers can be treated as a typical scattering problem, with \eqref{EincFE} as the incident field.

An important feature of \eqref{EincFE} is that the incident field is entirely evanescent (the asymptotic decay of the special function $K_n$ is given by $e^{-kr}/kr$ in the far field). This is expected on physical grounds, as an electron moving at constant velocity cannot radiate. Once a scattering body is brought close to the electron beam, however, the situation changes: the evanescent incident field can excite modes in the scatterer that couple to far-field radiation. (Physically, the electromagnetic-field-mediated interaction of the electron beam with the scatterer can lead to deceleration and therefore radiation.) The radiated power can be computed by an LDOS-like expression, $\frac{1}{2} \Re \int \Jv^* \cdot \Ev$, where $\Jv$ is the free-electron current density, but the bound techniques developed below for scattering bodies are most easily applied to the polarization fields $\Pv$ within the scatterer, so we prefer an equivalent expression in terms of $\Pv$. One option would be a linear combination of a direct-radiation term with a scatterer-interaction-radiation term, but the evanescent-only nature of the incident field implies that the direct-radiation term is zero. Instead, the only power lost by the electron beam is that which is extinguished by the scatterer, into absorption losses or far-field radiation. As we discuss more thoroughly in \secref{global}, the extinction of a scattering body $V$ is given by
\begin{align}
    \Pext = \frac{\omega}{2} \Im \int_V \Einc^*(\xv) \cdot \Pv(\xv) \dxv,
    \label{eq:Pext}
\end{align}
which we will use to analyze the free-electron loss\index{free-electron radiation}, as $P_{\rm loss} = \Pext$.

When the beam passes by the scatterer without intersecting it, the resulting radiation is referred to as {\bf Smith--Purcell radiation}\index{Smith--Purcell radiation}. When the beam passes through the scatterer, causing radiation, it is referred to as {\bf transition radiation}. And when the beam radiates while propagating \emph{inside} a refractive medium (within which the modified speed of light can be smaller than the electron speed), it is referred to as {\bf Cherenkov radiation}. The Smith--Purcell process resides squarely in the realm of near-field\index{near field} electromagnetism.

\subsection{CDOS}\index{cross density of states (CDOS)}
\label{sec:CDOS}
In \secref{LDOS}, we showed that the power radiated by a single dipole at position $\xv$ is proportional to the LDOS at that point, which itself is proportional to $\Im \GG(\xv,\xv)$. Consider now the power radiated by \emph{two} dipoles, $\pv_1$ and $\pv_2$, at positions $\xv_1$ and $\xv_2$, for a total dipole density of $\Pv(\xv) = \pv_1 \delta(\xv - \xv_1) + \pv_2 \delta(\xv-\xv_2)$. The power they jointly radiate is given by
\begin{align}
    P_{\rm rad} &= \frac{\omega}{2} \int_V \int_{V'} \Pv(\xv) \Im \GG(\xv,\xv') \Pv(\xv') \,{\rm d}\xv \,{\rm d}\xv' \nonumber \\
                &= \frac{\omega}{2} \left\{ \pv_1^\dagger \left[\Im \GG(\xv_1,\xv_1)\right] \pv_1 + \pv_2^\dagger \left[\Im \GG(\xv_2,\xv_2)\right] \pv_2 \right. \nonumber \\
                & \left. + \pv_1^\dagger \left[\Im \GG(\xv_1,\xv_2)\right] \pv_2 + \pv_2^\dagger \left[\Im \GG(\xv_2,\xv_1)\right] \pv_1 \right\}.
\end{align}
The first two terms are the powers radiated by the two dipoles in isolation (or when incoherently excited); the second pair of terms is the positive or negative contribution that arises for constructive or destructive (coherent) interference between the two dipoles. For reciprocal media (of arbitrary patterning), the third and fourth terms are complex-conjugates of each other, such that we can just consider one of them (say, the third term) in determining the two-dipole interference. By analogy with \eqref{LDOSe}, we can define a \textbf{cross density of states (CDOS)}\index{cross density of states (CDOS)} by the expression:
\begin{align}
    {\rm CDOS}_{ij}(\omega,\xv_1,\xv_2) = \frac{1}{\pi\omega} \Im \GG_{ij}(\xv_1,\xv_2) ,
    \label{eq:CDOS}
\end{align}
which differs from \citeasnoun{Caze2013} only by the absence of a 2 in the prefactor. The sign of the CDOS indicates the sign of the interference term, while its magnitude is a field-correlation strength between the two points of interest in a given electromagnetic environment. The amplification of emission that can occur when the sign is positive is an example of superradiance, while the reduction of emission when the sign is negative is an example of subradiance, in each case mediated by the local CDOS~\cite{Carminati2022}. Because the CDOS is the off-diagonal part of a positive-definite matrix, it is straightforward to show that its magnitude is bounded above by the square root of the product of the diagonal terms in the matrix, i.e., the local densities of states of the two dipoles in isolation~\cite{Canaguier-Durand2019}.

In systems that are closed, or approximately closed, there is another interesting interpretation of the CDOS~\cite{Caze2013,Canaguier-Durand2019}. Just as the LDOS can be interpreted as a local modal density, the CDOS can be intepreted as a local modal \emph{connectivity}---it is a measure of spatial coherence between two points. In \citeasnoun{Caze2013}, it was shown the one can compute local coherence lengths from spatial integrals of the CDOS. From these local coherence lengths, it was unambiguously demonstrated that ``spatial squeezing'' of eigenmodes occurs in systems of disordered plasmonic\index{plasmons} nanoparticles. This plausibly explains suprising experimental results when probing the local response of such disordered films~\cite{Krachmalnicoff2010}, showing the value of CDOS as an independent concept from LDOS.

There are two other areas in which CDOS emerges as a key metric: Forster energy transfer~\cite{dung_knoll_welsch_2002, martin-cano_2010, gonzaga-galeana_zurita-sanchez_2013} and quantum entanglement and super-radiative coupling between qubits~\cite{kastel_fleischhauer_2005, kastel_laser_2005, dzsotjan_sorensen_fleischhauer_2010, martin-cano_2011, gonzalez-tudela_martin-cano_2011}. The general idea in each case is a dipole $\pv_1$ transferring energy to a second dipole $\pv_2$. In this scenario, $\pv_1$ and $\pv_2$ are considered \emph{fixed}. By Poynting's theorem, the energy flux into a small bounding surface of $\pv_2$, for a field $\Ev_1$ generated by $\pv_1$, is
\begin{align}
    \frac{\omega}{2} \Im \left[ \pv_2^\dagger \Ev_1(\xv_2) \right] = \frac{\omega}{2} \Im \left[ \pv_2^\dagger \GG(\xv_2,\xv_1) \pv_1 \right],
\end{align}
which is a form of the CDOS\index{cross density of states (CDOS)}. The fixed nature of the second dipole, $\pv_2$, is crucial for the CDOS metric to be the correct one. If the second dipole is \emph{induced by the field emanating from the first dipole}, then $\pv_2 = \alpha_2 \Ev_1(\xv_2)$, and the correct energy-transer expression would be the imaginary part of the polarizability multiplied by the squared absolute value of the Green's function.

\subsection{Surface-enhanced Raman scattering (SERS)}
\label{sec:SERS}
\textbf{Surface-enhanced Raman scattering}\index{surface-enhanced Raman scattering (SERS)} is a technique whereby molecules are excited by a pump field, subsequently emitting Stokes- (or anti-Stokes-) shifted radiation that can be used for imaging or identification~\cite{Otto1992,Nie1997,Kneipp1997,Kneipp2002}. The small cross-sections of most chemical molecules results in very low pump and emission efficiencies in conventional Raman spectroscopy~\cite{Long1977}, but one can engineer the near-field\index{near field} environment to enhance both the concentration of the pump field as well as the emission rate. Efficiency improvements of up to 12 orders of magnitude have been demonstrated, enabling single-molecule detection and a variety of applications.

SERS is a nonlinear process, in which a single dipolar molecular sees both a pump enhancement as well as a spontaneous-emission\index{spontaneous emission} enhancement. A key insight for understanding SERS is that the weakness of the nonlinearities of the individual molecules means that the nonlinear process can be treated as the \emph{composition} of linear processes, in which the pump first enhances the excited-population densities (or, classically, the dipole amplitudes), and then the spontaneous-emission\index{spontaneous emission} enhancements can be treated as a second step, essentially independent of the first.

We can write the key metric of SERS by considering these two steps in sequence, following a procedure outlined in \citeasnoun{Michon2019}. First, an illumination field at frequency $\omega_0$ impinges upon the molecule and its environment; in tandem, a total field of $\Ev_{\omega_0}(\xv_0)$ is generated at the molecule. The Raman process generates a dipole moment at frequency $\omega_1$ given by
\begin{align}
    \pv_{\omega_1} = \alphat_{\rm Raman} \Ev_{\omega_0}(\xv_0)
\end{align}
where $\alphat_{\rm Raman}$ is the molecular polarizability. Next, the power radiated at $\omega_1$ by this dipole is given, per \eqref{PRadDip}, by
\begin{align}
    P_{\textrm{rad},\omega_1} = \pv_{\omega_1}^\dagger \left[\Im \GG_{\omega_1}(\xv_0,\xv_0)\right] \pv_{\omega_1}.
\end{align}
Hence we see that there are two opportunities for amplification of SERS: concentrating the incoming field $\Ev_{\omega_0}$ that determines the dipole amplitude, and enhancing the outgoing radiation by maximizing the LDOS\index{local density of states (LDOS)}, proportional to $\Im \GG_{\omega_1}(\xv_0,\xv_0)$, at the location of the dipole. To separate the two contributions, we can write the dipole moment as $\pv = \| \alphat \Ev \| \left( \alphat\Ev / \|\alphat \Ev\| \right)$, i.e., an amplitude multiplied by a unit vector. If we denote the unit vector as $\hat{\pv}_{\omega_1}$, then we can write 
\begin{align}
    P_{\textrm{rad},\omega_1} = \|\alphat_{\rm Raman} \Ev_{\omega_0}\|^2 \hat{\pv}_{\omega_1}^\dagger \left[\Im \GG_{\omega_1}(\xv_0,\xv_0)\right] \hat{\pv}_{\omega_1},
\end{align}
where now the first term encapsulates $\omega_0$-frequency concentration, and the second term encapsulates $\omega_1$-frequency LDOS-enhancement\index{local density of states (LDOS)}. Straightforward arguments lead to a net SERS\index{surface-enhanced Raman scattering (SERS)} enhancement, relative to a base rate $P_0$ without any nearby surface, given by
\begin{align}
    \frac{P_{\textrm{rad},\omega_1}}{P_0} = \left(\frac{\|\alphat_{\rm Raman}\Ev_{\omega_0}\|^2}{\|\alphat_{\rm Raman}\|^2 \|\Ev_{\textrm{inc},\omega_0}\|^2}\right) \left(\frac{ \rho_{\hat{\pv},\omega_1}}{\rho_{0,\omega_1}}\right),
    \label{eq:SERSEnh}
\end{align}
where $\|\alphat\|$ refers to the induced matrix norm of $\alphat$, $\rho_{\hat{\pv},\omega_1}$ is the $\omega_1$-frequency LDOS for a $\hat{\pv}$-polarized dipole, and $\rho_{0,\omega_1}$ in this expression is the background $\omega_1$-frequency LDOS of a $\hat{\pv}$-polarized dipole (not the typical summation over all polarizations). The two parenthetical terms in \eqref{SERSEnh} must both be bounded to identify fundamental limits to SERS enhancements.

\subsection{Near-field radiative heat transfer}
\label{sec:NFRHT}
The warming of the cold earth by the hot sun is mediated by radiative transfer, i.e., photons radiated from the sun to the earth. The maximum rate at which such a process could occur is of course given by the {\bf blackbody} rate, which is determined only by the solid angle subtended by the earth from the sun (or vice versa). Determination of this blackbody rate requires no knowledge of multiple-scattering processes between the two bodies. In the far field, the only ``channels'' (carriers of power into and out of a scattering region) are propagating-wave channels; by Kirchhoff's Law~\cite{Lienhard2011}, one need only know the absorption or emission rates of the two bodies in isolation to know their maximum radiative-exchange rate. A more general viewpoint of far-field radiation, via the idea of communication channels\index{communication channels}, is discussed in \secref{channel}.

It has been known for 75 years~\cite{Polder1971,Rytov1988} that two bodies separated by less than a thermal wavelength can exchange radiative heat at significantly larger rates than their far-field counterparts. Once in the near field\index{near field}, the bodies can exchange photons not only through radiative channels but additionally evanescent channels; moreoever, as the separation distance $d$ is reduced, the number of evanescent channels that can be accessed increases dramatically, scaling as $1/d^2$. These channels can be accessed via any mechanism that produces strong near fields\index{near field}. Polaritonic surface waves\index{polaritons}, via either plasmons\index{plasmons} or phonon--polariton\index{polaritons} materials, are a natural choice, and hyperbolic metamaterials (whose strongest effect is not surface waves but instead high-wavenumber bulk modes with nonzero evanescent tails) can provide similar performance~\cite{Biehs2013,Miller2014b}. Photonic crystals can also support surface waves, but the confinement of those waves is typically related to the size of their bandgap~\cite{Joannopoulos2011}, thereby scaling with frequency, yielding surface waves with significantly less confinement than their metallic counterparts.

The complexity of near-field\index{near field} radiative heat transfer (NFRHT)\index{near-field radiative heat transfer (NFRHT)} is daunting, both experimentally and theoretically. The first experimental demonstrations of enhancements in NFRHT via near-field\index{near field} coupling were not achieved until until the 2000's~\cite{Shen2009,Rousseau2009,Song2015}, many decades after the original predictions~\cite{Polder1971,Rytov1988}, and measurements in the extreme near field\index{near field} were not achieved until 2015~\cite{Kim2015}. There are a number of technical hurdles to experimental measurements, especially maintaining consistent, nanometer-scale gap separations over large-scale device diameters, while simultaneously measuring miniscule heat currents~\cite{Kim2015}.

The theoretical challenge has been no less severe. NFRHT\index{near-field radiative heat transfer (NFRHT)} involves rapidly decaying near fields\index{near field} (requiring high resolution), typically over large-area surfaces (requiring a large simulation region), for spatially incoherent and broadband thermal sources (such that the equivalent of very many simulations are needed). The computational complexity of this endeavor has limited the analysis of NFRHT\index{near-field radiative heat transfer (NFRHT)} almost exclusively to high-symmetry structures (planar/spherical bodies, metamaterials, etc.)~\cite{Loomis1994,Pendry1999,Joulain2005,Ben-Abdallah2009,Biehs2010,Kruger2011}, small resonators~\cite{Joulain2005,Mulet2001}, two-dimensional systems~\cite{Rodriguez2011} and the like. We review the planar-body interaction, which is informative, while emphasizing the need (and opportunity) for new theoretical tools to understand what is possible when exchanging radiative heat in the near field\index{near field}.

Consider two near-field\index{near field} bodies with temperatures $T_1$ and $T_2$, respectively. By the fluctuation--dissipation theorem, the incoherent currents in body 1, $\Jv_1$, have ensemble averages (denoted $\langle \rangle$) given by~\cite{Joulain2005}
\begin{align}
    \langle \Jv_1(\xv,\omega) \Jv_1^\dagger(\xv',\omega) \rangle = \frac{4\varepsilon_0 \omega}{\pi} \Im \left[\chi_1(\xv,\omega)\right] \Theta(\omega,T_1) \delta(\xv-\xv') \mathcal{I},
    \label{eq:FDT}
\end{align}
where $\chi_1(\xv,\omega)$ is the material susceptibility of body 1, $\mathcal{I}$ is the 3$\times$3 identity matrix, and $\Theta(\omega,T)$ is the Planck distribution,
\begin{align}
    \Theta(\omega,T) = \frac{\hbar\omega}{e^{\hbar\omega/kT} - 1}.
\end{align}
These currents radiate to body 2, at each frequency $\omega$, at a rate that we denote $\Phi_{21}(\omega)$. The rate $\Phi_{21}(\omega)$ is given by the ensemble average of the flux into body 2, i.e. $\langle -\frac{1}{2}\Re\int_{S_2} \Ev \times \Hv^* \cdot \hat{\vect{n}} \rangle$, where $S_2$ is a bounding surface of $V_2$, $\hat{\vect{n}}$ is the outward normal, and the field sources are given by \eqref{FDT}, except without the Planck function. The Planck function is separated so that $\Phi_{21}(\omega)$ is independent of temperature and depends only on the electromagnetic environment. Then the radiative heat transfer rate into 2 from currents in 1, denoted $H_{21}$, is given by
\begin{align}
    H_{21} = \int \Phi_{21}(\omega) \Theta(\omega,T_1) \dw.
\end{align}
Similarly, the rate of transfer from body 2 to body 1, $H_{12}$, is given by
\begin{align}
    H_{12} = \int \Phi_{12}(\omega) \Theta(\omega,T_2) \dw,
\end{align}
and the net transfer rate is the difference between the two. For reciprocal bodies, the rates $\Phi_{12}(\omega)$ and $\Phi_{21}(\omega)$ are always equal (by exchanging the source and ``measurement'' locations), but this is also true more generally: for \emph{two} bodies exchanging radiative heat in the near field\index{near field}, $\Phi_{12}(\omega)$ and $\Phi_{21}(\omega)$ must be equal, or else one could have net energy exchange with both bodies at equal temperatures, in violation of the second law of thermodynamics. Note that if three bodies are present, or either body radiates significant amounts of energy into the far field, this relation need not hold in nonreciprocal systems, and indeed ``persistent currents'' have been predicted in three-body systems in the near field\index{near field}~\cite{Zhu2016}. Throughout this chapter we will focus on the prototypical two-body case, so we can take
\begin{align}
    \Phi_{12}(\omega) = \Phi_{21}(\omega) = \Phi(\omega),
\end{align}
without assuming reciprocity. Hence the \emph{net} NFRHT\index{near-field radiative heat transfer (NFRHT)} rate between the two bodies is given by
\begin{align}
    H_{2\leftarrow 1} = \int \Phi(\omega) \left[ \Theta(\omega,T_1) - \Theta(\omega,T_2) \right] \dw.
    \label{eq:H2r1}
\end{align}
Often, it is illuminating to reduce the problem to a single temperature $T$ and study the differential heat transfer for a temperature differential $\Delta T$. The net heat exchange divided by this temperature differential is the {\bf heat transfer coefficient}\index{heat transfer coefficient (HTC)}, or HTC, which is given by \eqref{H2r1}, except the temperature difference is replaced by a single derivative of $\Theta(\omega,T)$ with respect to temperature:
\begin{align}
    {\rm HTC} = \int \Phi(\omega) \frac{\partial \Theta(\omega,T)}{\partial T} \dw.
    \label{eq:HTC}
\end{align}
Hence, the quantity $\Phi(\omega)$ is the designable quantity in NFRHT\index{near-field radiative heat transfer (NFRHT)}, and is the focus of the NFRHT\index{near-field radiative heat transfer (NFRHT)} bounds appearing across \secref{NearFieldLimits}.

\subsection{Mode volume}
\label{sec:modev}
Finally, we turn to a unique near-field\index{near field} quantity: mode volume\index{mode volume}. Intuitively, \textbf{mode volume}\index{mode volume} encapsulates an ``amount of space'' occupied by an electromagnetic mode. Obviously, defining the volume of a continuous density is necessarily subjective. But we can develop an intuitive approach to the common volume definition. The energy density of a mode $m$ at any point $\xv$ is proportional to $\varepsilon(\xv) |\Ev_m(\xv)|^2$. If the maximum energy density occurs at a point $\xv_0$, we can define the volume of the mode as follows: let us redistribute the energy into a binary pattern in which at every point in space it can only take the values 0 or $\varepsilon(\xv_0) |\Ev_m(\xv_0)|^2$. Let us also require that the total energy of the mode not change in this binarization, i.e., $\int \varepsilon(\xv) |\Ev(\xv)|^2$ remains fixed. Then the corresponding redistributed field will occupy the volume:
\begin{align}
    V_m = \frac{\int \varepsilon(\xv) |\Ev_m(\xv)|^2}{\varepsilon(\xv_0)|\Ev_m(\xv_0)|^2}.
    \label{eq:Vm}
\end{align}
Typical modes of interest, which have strong field concentration and Gaussian- or Lorentzian-like energy decay, are well-suited to such an interpretation.

More rigorously, per \eqref{lorentz}, the modal field intensity is the quantity that determines the interaction of a dipole with a specific mode, and the contribution of that mode to the spontaneous emission\index{spontaneous emission} of the dipole. Then an alternative interpretation of the quantity in \eqref{Vm} is that the numerator can be taken to be 1, for a normalized mode, and the denominator is the relevant coupling term in the Hamiltonian that is to be maximized. This alternative approach explains why a common mathematical objective is to minimize the expression in \eqref{Vm}, without reference to any physical concept of volume.

A critical question around mode volume\index{mode volume} is whether such a concept is even valid. For closed (or periodic) systems with nondispersive, real-valued permittivities, the Maxwell operator is Hermitian, and there is an orthogonal basis of modal fields that can be orthonormalized. Dispersion in the material systems makes the eigenproblem nonlinear, but for Drude--Lorentz-like dispersions, one can introduce auxliary variables, and in this higher-dimensional space there is again a linear, Hermitian eigenproblem~\cite{Raman2010}. But once losses are introduced, either through open boundary conditions or material dissipation, the operator is no longer Hermitian, and the modes cannot be orthonormalized with an energy-related inner product~\cite{Lalanne2018}. Instead, one must work with \emph{quasinormal modes} (QNMs)\index{quasinormal modes (QNMs)}, for which two issues arise. 

If material losses are the dominant loss mechanism, as is typical in plasmonics\index{plasmons}, then the key new subtlety often is the modification of orthogonality: the modes are orthogonal in an \emph{unconjugated} ``inner product'' (e.g. $\int \varepsilon \Ev_1 \cdot \Ev_2$ instead of $\int \varepsilon \Ev_1^* \cdot \Ev_2$), which then replaces the standard conjugated inner product in modal expansions such as \eqref{lorentz}. While this is mathematically convenient, it can stymie our typical intuition. A beautiful example is demonstrated in \citeasnoun{Sauvan2013}. There, it is shown that the spontaneous emission\index{spontaneous emission} near a two-resonator antenna can be dominated by two QNMs\index{quasinormal modes (QNMs)}, as expected. However, if one tries to attribute individual contributions from each QNM, one of the QNMs appears to contribute \emph{negative} spontaneous emission\index{spontaneous emission}. This is attributable to the modified inner product: modes that are orthogonal in the unconjugated inner product are not orthogonal in an energy inner product, and their contributions to a positive energy flow (such as spontaneous emission\index{spontaneous emission}) are invariably linked; one can no longer separate a power quantity such as LDOS\index{local density of states (LDOS)} into individual contributions from constituent modes. Ultimately, one can define mode volume\index{mode volume} as a complex-valued quantity~\cite{Sauvan2013}, in which case it no longer becomes an independent quantity of interest to minimize or maximize, but rather an ingredient for other scattering quantities of interest.

If radiation losses are the dominant loss mechanism, one faces a hurdle even before orthogonality: just \emph{normalizing} the modal fields becomes tricky. If the modal fields eventually radiate in free space, they will asymptotically scale as $e^{ik_mr}/r$, where $k_m=\omega_m/c$ is the wavenumber of the mode and $r$ is a distance from the scatterer. But the losses to radiation transform the resonant eigenvalues to poles in the lower-half of the complex-frequency plane, i.e., $\omega_m \rightarrow \omega_m^{(r)} - i\omega_m^{(i)}$, where $\omega_m^{(i)} > 0$. Hence the modal fields grow exponentially, $\sim e^{\omega_{m}^{(i)} r}$, such that any integrals of the form $\int \Ev^2$ or $\int |\Ev|^2$ diverge. There are a few resolutions to this issue~\cite{Lalanne2018}. Perhaps the simplest is to use computational perfectly matched layers (PMLs) to confine the fields to a finite region. Then, for any accurate discretization of the Maxwell operator, one is simply left with a finite-sized, non-Hermitian matrix, whose eigenvectors will generically be orthonormalizable under the unconjugated inner product. (Exceptions to this occur at aptly named \emph{exceptional points}\index{exceptional points}, where modes coalesce, and one needs Jordan vectors to complete the basis~\cite{Kato2013,Brody2013}.) The orthonormalization of these modes faces the same interpretation issues discussed above in the plasmonic\index{plasmons} case, and there is one further difficulty: sometimes important contributions to energy expression can come from fields that \emph{primarily reside in the PML region}. It is difficult to attribute physical intuition or meaning to such contributions.

In \secref{modevbounds}, where we develop bounds for mode volume\index{mode volume}, we will only deal with cases of lossless dielectric materials, and we assume the quality factors are sufficiently high that the system is approximately closed. This is the limit in which the mode volume\index{mode volume} as defined by \eqref{Vm} is exactly the quantity that enters the LDOS\index{local density of states (LDOS)} expression of \eqref{lorentz}, which is typically the underlying goal of minimizing mode volume\index{mode volume} in the first place. In scenarios where one must use quasinormal modes\index{quasinormal modes (QNMs)}, it is probably better to eschew them altogether (if one wants a bound), and to instead work directly with the scattering quantity (e.g. LDOS) of interest.

\section{Analytical and computational bound approaches}
\label{sec:BoundApproaches}
Across many areas of science and technology, ``fundamental limits'' or ``bounds'' play an important role in technological selection, theoretical understanding, and optimal design. Examples abound:
\begin{itemize}
    \item The {\bf Shockley--Queisser limits}\index{Shockley--Queisser limit} for solar-cell energy conversion efficiency. Originally developed for single-cell, all-angle solar absorption and energy conversion~\cite{Shockley1961}, the basic framework they developed identifies two required loss mechanisms in any solar cell: radiation back to the sun (at the open circuit condition~\cite{Miller2012}), and thermalization losses in the establishment of quasi-Fermi levels in each band. Almost any proposed solar-energy-conversion technique must be put through a Shockley--Queisser analysis to earn serious consideration as a technology.
    \item The {\bf Yablonovitch $4n^2$ limit}\index{Yablonovitch limit}, for the maximum broadband, all-angle absorption enhancement in any optically thick material~\cite{Yablonovitch1982}. The factor $4n^2$, for a refractive index $n$, arises from the density-of-states enhancement in a high-index material, a 2X enhancement from mirrors on the rear surface, and a 2X enhancement from the reorientation of mostly-vertical rays into random angles.
    \item The {\bf Wheeler--Chu limit}\index{Wheeler--Chu limit} to antenna quality factor, $Q$~\cite{Wheeler1947,Chu1948}. It is difficult for a subwavelength antenna (such as a cell-phone antenna) to operate over a wide bandwidth, and the Wheeler--Chu (sometimes Harrington is also given credit~\cite{Harrington1960}) limit imposes a bound on the maximum operational bandwidth. Most state-of-the-art antenna designs operate very close to the Wheeler--Chu limit~\cite{Sievenpiper2012}.
    \item The {\bf Bergman--Milton bounds}\index{Bergman--Milton bounds} on the effective properties of a composite material~\cite{Bergman1980,Milton1980,Bergman1981,Milton1981,Milton1981a,Kern2020}.
    \item The {\bf Abbe diffraction limit}\index{Abbe diffraction limit} on the maximum focusing of an optical beam. This limit can be circumvented in the near field\index{near field}~\cite{Fang2005,Merlin2007}, or even in the far field if one is willing to tolerate side lobes~\cite{McCutchen1967,Stelzer2002,Zheludev2008,Slepian1961,Landau1961,Ferreira2006,Shim2020}.
    \item The {\bf Shannon bounds}\index{Shannon bounds}~\cite{Shannon1949}, a foundational idea in information theory~\cite{Cover1999}.
\end{itemize}
Many of these examples involve electromagnetism, but typically only for noninteracting waves and simplified physical regimes. The Yablonovitch $4n^2$ limit applies in geometric (ray) optics, the Wheeler--Chu limit only arises in highly subwavelength structures, and the diffraction limit applies only to free space (or homogeneous-medium) propagation. Is it possible to create an analogous theoretical framework for the full Maxwell equations, identifying fundamental spectral response bounds while accounting for the exceptional points\index{exceptional points}~\cite{Heiss2012,Miri2019}, speckle patterns~\cite{Bender2019}, bound states in the continuum~\cite{Hsu2016}, and other exotic phenomena permitted by the wave equation? A flurry of work over the past decade suggests that in many scenarios, the answer should be ``yes.'' In the following subsections we outline the key new ideas that have been developed.

\subsection{Global conservation laws}
\label{sec:global}
One approach particularly well-suited to formulating bounds is to replace the complexity of the full Maxwell-equation design constraints with a single constraint that encodes some type of conservation law. The Yablonovitch limit, discussed in the previous section, offers a powerful example: to identify maximum absorption enhancement in a geometric-optics setting, one can replace the complexity of ray-tracing dynamics with a single density-of-states constraint. Unfortunately, one cannot extend such density-of-states arguments to full-Maxwell and near-field\index{near field} settings, but other types of ``conservation laws'' can be identified. A {\bf global conservation law}\index{global conservation laws} that has been particularly fruitful for nanophotonics is the \textbf{optical theorem}\index{optical theorem}. The optical theorem\index{optical theorem}~\cite{Newton1976,Jackson1999,Lytle2005} is a statement of global power conservation: the total power extinguished from an incident beam by a scattering body (or bodies) equals the sum of the powers scattered and absorbed by that body. Writing the extinguished, scattered, and absorbed powers as $\Pext$, $\Pscat$, and $\Pabs$, respectively, the optical theorem\index{optical theorem} can be expressed as
\begin{align}
    \Pext = \Pscat + \Pabs.
\end{align}
Conventionally, the optical theorem\index{optical theorem} is specified in terms of the far-field scattering amplitudes of a scattering body~\cite{Newton1976}, in which case the extinction is shown to be directly proportional to the imaginary part of the forward-scattering amplitude. This expression can be interpreted as a mathematical statement of the physical intuition that the total power taken from an incident beam can be detected in the phase and amplitude of its shadow. The analysis does not have to be done in the far field; another common version is to relate the extinguished-, scattered-, and absorbed-power fluxes via surface integrals of the relevant Poynting fluxes~\cite{Jackson1999}. Still one more version of the optical theorem\index{optical theorem}, and the one that turns out to be most useful for wide-ranging bound applications, is to the use the divergence theorem to relate the surface fluxes to the fields within the volume of the scatterer, and write all powers in terms of the polarization currents and fields induced in those scatterers~\cite{Lytle2005}. As we briefly alluded to in the discussion of free-electron radiation\index{free-electron radiation} in \secref{FER}, the work done by a field $\Ev$ on a polarization field $\Pv$ in a volume $V$ is given by $\left(\frac{\omega}{2}\right) \Im \int_V \Ev^* \cdot \Pv = \left(\frac{\omega}{2}\right) \int_V \Pv^* \left[ \Im \chi / |\chi|^2 \right] \Pv$, where $\chi$ is the material susceptibilty. (We assume throughout scalar, electric material susceptibilities $\chi$. The generalizations to magnetic, anisotropic, and bianisotropic materials are straightforward in every case.) Extinction is the work done by the incident field on the induced polarization field, scattered power is the work done by that polarization field on the scattered fields $\Escat$, and absorbed power is the work done by the total field on the polarization field. Hence the optical theorem\index{optical theorem} reads:
\begin{align}
    \Im \int_V \Einc^*(\xv) \cdot \Pv(\xv) \dxv = \Im &\int_V \int_V \Pv^*(\xv) \cdot \GG_0(\xv,\xv') \Pv(\xv') \,{\rm d}\xv \,{\rm d}\xv' \nonumber \\
                                                      &+ \int_V \Pv^*(\xv) \cdot \frac{\Im \chi(\xv)}{|\chi(\xv)|^2} \Pv(\xv) \dxv,
    \label{eq:optthmint}
\end{align}
where we have substituted $\Escat(\xv) = \int_V \GG_0(\xv,\xv') \Pv(\xv') \dxv'$ for the scattered field and dropped the constant factor $(\omega/2)$ preceding every integral. \Eqref{optthmint} relates extinction on the left-hand side to the sum of scattered and absorbed powers on the right-hand side. For intuition and compactness, it is helpful to rewrite equations like \eqref{optthmint} in a matrix/vector form. We can assume any arbitrarily high-resolution discretization in which $\Pv(\xv)$ becomes a vector $\pv$, the integral operator $\int_V \GG(\xv,\xv') \dxv'$ becomes a matrix $\GG_0$, and integrals of the conjugate of a field $\vect{a}(\xv)$ with another $\vect{b}(\xv)$ are replaced with vector inner products $\vect{a}^\dagger \vect{b}$. It is also helpful to define a material parameter $\xi(\xv) = -1/\chi(\xv)$, and a corresponding (diagonal) matrix $\xi = -\chi^{-1}$. With these notational changes, \eqref{optthmint} can be re-written
\begin{align}
    \Im \left(\einc^\dagger \pv\right) = \pv^\dagger \left[\Im \GG_0 + \Im \xi\right] \pv.
    \label{eq:optthm}
\end{align}
This is the vectorized version of the optical theorem\index{optical theorem}, and it illuminates some of the mathematical structure embedded in this particular version of power conservation. The left-hand side is a linear function of the polarization field $\pv$, while the right-hand side is a quadratic function. Moreover, in passive systems\index{passivity} the absorbed and scattered powers are nonnegative quantities. This nonnegativity is embedded in the matrices (operators) $\Im \GG_0$ and $\Im\xi$, both of which are positive semidefinite (denoted by ``$\geq 0$'') in passive systems\index{passivity}:
\begin{align}
    \Im \GG_0 &\geq 0, \\
    \Im \xi &\geq 0.
\end{align}
The positive semidefinite nature of these matrices implies that the right-hand side of \eqref{optthm} is a convex quadratic functional of $\pv$. Hence \eqref{optthm} can be interpreted as an ellipsoid (as opposed to a hyperboloid) in the high-dimensional space occupied by $\pv$. 

A key feature of \eqref{optthm}, and the conservation laws to follow, is that it is ``\textbf{domain oblivious}''\index{domain oblivious}~\cite{Kuang2020}. Suppose we enforce that constraint on a high-symmetry domain, such as a sphere or half-space, where the operator $\GG_0$ might be easy to construct. Of course, enforcing \eqref{optthm} will enforce power conservation on the sphere itself. But it \emph{also enforces power conservation on all sub-domains of the sphere}. This is not obvious--the operator $\GG_0$ is different for every choice of domain and range, and once we have chosen a sphere for both, it seems that we are stuck with only the sphere domain. The key, however, is the appearance of $\pv$ in each term of \eqref{optthm}, and twice on the right-hand side. To enforce \eqref{optthm} on a smaller sub-domain, then instead of changing the domain and range of the operator, we can instead enforce the polarization $\pv$ to be zero at each point outside the sub-domain but inside the enclosing domain. On the right-hand side, this effectively changes both the domain and range of $\GG_0$, while on the left-hand side, it nulls any extinction contribution from outside the sub-domain. Hence, the conservation law of \eqref{optthm}, and all of the volume-integral-based conservation laws to follow, is domain oblivious\index{domain oblivious}.

Power conservation via the optical theorem\index{optical theorem} has led to a surprisingly wide array of bounds and fundamental limits in electromagnetic systems. The key idea is to drop the full Maxwell-equation constraint that is implicit in any design problem, and replace it with only the power-conservation expression of \eqref{optthm}. Even with just this single constraint, surprisingly good bounds can be attained. As an example, consider systems where absorptive losses are more important than radiation/scattering losses. In such systems, we can drop the $\Im\GG_0$ term in the optical theorem\index{optical theorem} of \eqref{optthm}, and use its positivity to write a constraint that absorbed power be less than or equal to extinction:
\begin{align}
    \pv^\dagger \left( \Im \xi \right) \pv \leq \Im \left(\einc^\dagger \pv\right).
    \label{eq:optthmabs}
\end{align}
This constraint implies a bound on the strength of the polarization field, because the left-hand-side term is quadratic (and positive-definite) in $\pv$, while the right-hand side is linear in $\pv$. A few steps of variational calculus~\cite{Miller2016} can identify the largest polarization-field strength that can be induced in a scatterer:
\begin{align}
    \| \pv \|^2 = \pv^\dagger \pv = \int_V \left|\Pv(\xv)\right|^2 \dxv \leq \frac{\|\einc\|^2}{\Im \xi} = \frac{|\chi|^2}{\Im \chi} \int_V \left|\Einc(\xv)\right|^2 \dxv.
\end{align}
We have a first bound: in a lossy material, wherein $\Im \chi > 0$, there is a bound on the largest polariation currents that can be induced in a scatterer, based only on the material properties and the energy of the \emph{incident} wave in the scattering region. Polarization currents beyond this strength would have absorbed powers larger than their extinction, implying an unphysical negative scattered power.

Beyond the strength of the polarization field itself, one can use similar variational-calculus arguments to identify bounds on wide-ranging quantities: extinction, absorption, and scattering, in bulk materials~\cite{Miller2016}, 2D materials~\cite{Miller2017}, and lossy environments~\cite{Ivanenko2019,Nordebo2019}; high-radiative-efficiency scatterers~\cite{Yang2017}; and even near-field\index{near field} quantities such as local density of states~\cite{Miller2016,Michon2019}, near-field\index{near field} radiative heat transfer~\cite{Miller2015,Miller2017}, and Smith--Purcell radiation\index{Smith--Purcell radiation}~\cite{Yang2018}. As a canonical example, let us consider the extinction, absorption, and scattering cross-sections of a scattering body with volume $V$, susceptibility $\chi$, and a plane-wave incident field. Cross sections $\sigma_{\rm ext,abs,scat}$ are the relevant powers divided by the intensity of the incident wave; the corresponding bounds are
\begin{align}
    \frac{\sigma_{\rm abs,scat,ext}}{V} \leq \frac{\beta\omega}{c} \frac{|\chi|^2}{\Im \chi} \qquad \beta_{\rm abs,ext} = 1, \beta_{\rm scat} = \frac{1}{4}.
    \label{eq:csboundsabs}
\end{align}
Per-volume cross-sections are bounded above by the frequency of the incoming waves and the material susceptibilities. Plasmonic nanoparticles can approach these bounds~\cite{Miller2016,Miller2017,Miller2022}.

One subtletly that arises in the near field\index{near field} (whose bounds are discussed in depth in \secref{NearFieldLimits}) is which conservation laws to use. The absorption- and extinction-based constraint of \eqref{optthmabs} may not be ideal for local density of states, for example, as the power radiated by a dipole is not exactly the same as the power extinguished by a nearby scatterer. (There is a separate pathway for the dipole to radiate directly to the far field, and this radiation can destructively/constructively interfere with waves scattered by the scatterer.) The optical theorem\index{optical theorem} of \eqref{optthm} arises from equating fluxes through a surface surrounding the scatterer. Instead, in the near field\index{near field}, one can draw a surface around the dipolar source itself. Then one can identify new conservation laws, which now relate the total power radiated by the dipole (the LDOS\index{local density of states (LDOS)}) to the sum of power absorbed in the scatterer and power radiated to the far field.

In some systems, radiation losses are the limiting factor rather than absorption losses. Prominent examples include metals at low frequencies, and low-loss dielectrics. In these systems, the key component of the optical theorem\index{optical theorem} of \eqref{optthm} is the radiation-loss term with $\Im \GG_0$, not the absorption-loss term. Of course, absorption must be positive, so we can drop it and replace the optical theorem\index{optical theorem} with a second inequality version:
\begin{align}
    \pv^\dagger \left(\Im \GG_0\right) \pv \leq \Im \left( \einc^\dagger \pv \right).
    \label{eq:optthmscat}
\end{align}
Although the $\Im \GG_0$ matrix may appear daunting, we typically use high-symmetry volumes for our designable domains, and we can use analytical or semi-analytical forms of $\Im \GG_0$ in those domains. (Such usage does not restrict the validity of the bound to only the high-symmetry domain; as discussed above, this expression is domain oblivious\index{domain oblivious}.) One common high-symmetry domain is a sphere, in which case $\Im \GG_0$ can be written in a basis of vector spherical waves~\cite{Tsang2000,Molesky2020,Kuang2020b}. Application of this approach to the question of maximum cross-sections yields different bounds from the ones of \eqref{csboundsabs}. One must limit the number of spherical waves that can contribute to the scattering process; allowing only the first $N$ electric multipole leads to maximum cross-sections proportional to the square of the wavelength, $\lambda$:
\begin{align}
    \sigma_{\rm abs,scat,ext} \leq \frac{\beta\lambda^2}{\pi}\left(N^2+2N\right) \qquad \beta_{\rm scat,ext} = 1, \beta_{\rm abs} = \frac{1}{4},
    \label{eq:csboundsscat}
\end{align}
with double the value if the magnetic vector spherical waves can be equally excited. Note the different values of $\beta$ for absorption and scattering in the absorption-limited case of \eqref{csboundsabs} versus the radiation-limited case of \eqref{csboundsscat}. The different coefficients arise because of the different conditions under which maximum extinction occur. In an absorption-dominated system, arbitrarily small scattering is possible (in principle), such that the maximum for extinction and absorption coincide, while the scattered-power maximum requires a reduction in absorption relative to extinction and a $1/4$ coefficient to account for the matching that must occur. The opposite occurs in scattering-limited systems, where absorption can be arbitrarily small (in principle), the maximum for extinction and scattering coincide, and an extra factor of $1/4$ is introduced when absorption is to be maximized. The bound of \eqref{csboundsscat} was originally derived for antenna applications or spherically symmetry scatterers via long and/or restrictive arguments~\cite{Hamam2007,Kwon2009,Ruan2011,Liberal2014,Liberal2014b}; the single conservation law of \eqref{optthmscat} is sufficient to derive \eqref{csboundsscat} in quite general settings~\cite{Hugonin2015,Miroshnichenko2018}. (An interesting precursor to the global-conservation-law approach\index{global conservation laws} is \citeasnoun{Gustafsson2012}, which identifies metrics that intrinsically have bounded optima over polarization currents, even \emph{without} any constraints.)

Of course, in some settings both absorption and radiation losses will be important to capture what is possible, and the bounds of \eqreftwo{csboundsabs}{csboundsscat} may not be sufficient. It is possible to capture both loss mechanisms in a single bound by using the entirety of the optical theorem\index{optical theorem}, \eqref{optthm}, without dropping either term. This was first recognized in Refs.~\cite{Kuang2020b,Gustafsson2020,Molesky2020b}. \citeasnoun{Kuang2020b} used this approach to derive bounds on the thinnest possible perfect absorber. (Or, conversely, the maximum absorption of an arbitrarily patterned thin film with a given maximum thickness.) Cross-section bounds given in \citeasnoun{Kuang2020b,Gustafsson2020,Molesky2020b} are generalizations of the two bounds listed above, \eqreftwo{csboundsabs}{csboundsscat}, containing each as separate asymptotic limits. At normal incidence, one can derive a simple transcendental equation for the minimum thickness, $h_{\rm min}$, of a perfect absorber with material parameter $\xi = -1/\chi$:
\begin{align}
    h_{\rm min} = \left(\frac{2\lambda}{\pi}\right) \frac{\Im \xi(\omega)}{1 - \operatorname{sinc}^2\left(\omega h_{\rm min}/c\right)}.
\end{align}
This approach has been successfully applied to the identification of the minimum thickness of a metasurface reflector~\cite{Abdelrahman2022}.

Finally, at the global-conservation level\index{global conservation laws}, one can go one step further, as first recognized in Refs.~\cite{Gustafsson2020,Molesky2020b}. The optical theorem\index{optical theorem} of \eqref{optthm} represents the conservation of \emph{real} power across the volume of a scatterer, which can be understood as the conservation of the real part of the Poynting vector through any bounding surface. Additionally, the imaginary part of the Poynting vector, corresponding to what is known as \emph{reactive} power~\cite{Jackson1999}. The complex-valued version of the optical theorem\index{optical theorem} is essentially the same as \eqref{optthm} but without the imaginary part in any of the terms; a careful analysis leads to the generalized optical theorem\index{optical theorem}:
\begin{align}
    -\pv^\dagger \einc = \pv^\dagger \left[\GG_0 + \xi\right] \pv.
    \label{eq:optthmgen}
\end{align}
The real and imaginary parts of \eqref{optthmgen} now offer \emph{two} global conservation laws\index{global conservation laws} that must be satisfied in any scatterer. The real-power conservation law accounts for absorption- and radiation-loss pathways, while the reactive-power conservation law accounts for resonance conditions in real materials. The latter has been shown to be beneficial for tightening bounds in plasmonic\index{plasmons} materials that are relatively large (wavelength-scale sizes are quite large for plasmonic\index{plasmons} resonators) or which have very large negative real susceptibilities and/or very small imaginary susceptibilities~\cite{Molesky2020b}. This approach has been applied to bounds in cloaks~\cite{Jelinek2021} and focusing efficiency~\cite{Schab2022}. \Eqref{optthmgen} can be derived in one step from the volume-integral equation~\cite{Chew2008} (or Lippmann--Schwinger equation), which in this notation reads $\left[\GG_0 + \xi\right] \pv = -\einc$, simply by taking the inner product of that equation with $\pv$.

In this section we have seen that the optical theorem\index{optical theorem}, written over the volume polarization fields induced in a scatterer, offers a single (or two) global conservation laws\index{global conservation laws} that can be used to identify bounds in wide-ranging applications. In \secref{local} below we show that it is also a starting point for generating an infinite number of ``local'' conservation laws. First, however, we will explore an approach that is closely related to global conservation laws\index{global conservation laws}: so-called ``channel'' bounds.

\subsection{Channel bounds}
\label{sec:channel}
In this section, we explore another technique for identifying bounds to what is possible: decomposing power transfer into a set of independent or orthogonal power-carrying ``channels.'' Then the upper limits distill to the maximum power (or alternative objective) per channel multiplied by the number of possible channels.

A particularly elegant formulation of channels was proposed by D. A. B. Miller and colleagues in the early 2000's~\cite{miller1998spatial, miller2000communicating, piestun2000electromagnetic,miller2019waves}. Consider a transmitter region that wants to communicate (i.e. send information/energy) to a receiver region, and a vacuum (or background) Green's-function operator $\GG_0$ comprising the fields in the receiver from sources in the transmitter. How many communication channels\index{communication channels} are possible? There is a simple, rigorous mathematical answer to this question: if one decomposes the $\GG_0$ operator via a singular value decomposition (SVD)~\cite{Trefethen1997},
\begin{align}
    \GG_0 = \mathbb{U} \mathbb{S} \mathbb{V}^\dagger,
    \label{eq:GG0SVD}
\end{align}
then each pair of singular vectors forms an independent channel. The singular-value decomposition encodes orthogonality and normalization. For example, the first right singular value, which we can call $\vect{v}_1$, radiates \emph{only} to the first left singular vector $\vect{u}_1$ in the receiver region, and the strength of this connection is given exactly by the first singular value, which we can call $s_1$. This triplet $(\vect{v}_1,\vect{u}_1,s_1)$ mathematically define a \textbf{communication channel}\index{communication channels}, as are all the pairs in the SVD. There cannot be an infinite number of such channels with arbitrarily large strengths, as the channel strengths obey a simple sum rule\index{sum rules} related to the integral of the Green's function over the transmitter and receiver volumes:
\begin{align}
    \sum_i |s_i|^2 = \Tr\left( \mathbb{S}^\dagger \mathbb{S} \right) = \Tr\left(\GG_0^\dagger \GG_0 \right) = \int_{V_T}\int_{V_R} \left\| \GG_0(\xv_T,\xv_S) \right\|^2 \dxv_T \dxv_R.
\end{align}
One can define more granular bounds as well: for any transmitter/receiver regions enclosed within high-symmetry bounding domains, one can identify upper limits for each individual singular value~\cite{Kuang2022}. The singular values must decay exponentially in two-dimensional systems, whereas in three dimensions their decay can be sub-exponential. This SVD-based decomposition of \eqref{GG0SVD} implicitly uses a field-energy normalization; one can alternatively use power-transfer normalizations and arrive at related bounds for the communication strength between two volumes~\cite{ehrenborg2017fundamental,ehrenborg2020physical,ehrenborg2021capacity}. Each of these is a powerful approach for free-space communication systems such as MIMO~\cite{Gao2017,Asaad2018}. More generally, they capture a general truth about free-space propagation: it can always be decomposed into orthogonal, power-carrying channels.

In the near field\index{near field}, however, evanescent waves do \emph{not} offer an equivalent set of power-carrying channels. Evanescent waves obey different mathematical orthonormalization rules, which are consistent with the following fact: evanescent waves decaying (or growing) in one direction \emph{cannot} carry power; power can be transmitted only in the presence of oppositely directed evanescent waves~\cite{Johnson2002}. A prototypical example: a single interface can only exhibit total internal reflection alongside evanescent-wave excitation, whereas the introduction of second interface, and counter-propagating evanescent waves, can lead to the tunneling of power through a ``barrier.''

In lieu of the general SVD approach, in high-symmetry scenarios it is often possible to decompose power transfer in a high-symmetry basis. For example, a spherically symmetric scatterer preserves the quantum numbers of incoming vector spherical waves and cannot scatter into waves of different quantum numbers, which implies that each vector spherical wave comprises a ``channel'' for incoming and outgoing radiation. Similarly, in planar systems, the in-plane (parallel) wavevector $\vect{k}$ is a conserved quantity, in which case one can isolate the scattering process into each $\vect{k}$-dependent propagating and evanescent plane wave. One cannot define free-space evanescent-wave channels, per the orthonormalization discussion above, but a more complete analysis can lead to $\vect{k}$-dependent transfer coefficients that are readily interpretable as a channel-based power decomposition. We discuss the successful application of these ideas to near-field\index{near field} radiative heat transfer in \secref{NFRHTBounds}. A word of caution is important, however: the assumption of a high-symmetry structure dramatically limits the set of structures to which such bounds apply, and in many scenarios it has been found that the symmetry-independent approaches of global conservation laws\index{global conservation laws} (previous section) and local conservation laws\index{local conservation laws} (next section) yield both tighter \emph{and} more general bounds.

\subsection{Local conservation laws}
\label{sec:local}
In the global-conservation-law section of \secref{global}, we discussed that one or two conservation-of--power constraints is already sufficient for bounds in many scenarios of interest. Of course, one or two constraints cannot capture every objective of interest: if, for example, one wanted to know the largest average response over multiple incident fields, certainly more constraints are needed. Thankfully, it turns out that there is a systematic way to generate a large number of conservation-law constraints for any nanophotonic design problem of interest.

The key is to identify \textbf{\emph{local} conservation laws}\index{local conservation laws} that apply at every point within the scatterer~\cite{Kuang2020,Molesky2020}. These conservation laws can be ``built'' from a volume-integral formulation of the underlying governing dynamics, but we will use a more intuitive approach to develop them. The ``generalized optical theorem''\index{optical theorem} is written in \eqref{optthmgen} in vector/matrix notation; the equivalent integral expression is
\begin{align}
    \int_V \int_V \Pv^*(\xv) \GG_0(\xv,\xv') \Pv(\xv') \dxv \dxv' + \int_V \Pv^*(\xv) \xi(\xv) \Pv(\xv) \dxv = -\int_V \Pv^*(\xv) \Einc(\xv) \dxv.
\end{align}
To formulate local conservation laws\index{local conservation laws}, we simply recognize the following: for the first integral over the entire scatterer $V$ that appears in every term, we can replace $V$ with $V_{\xv}$, where $V_{\xv}$ is an infinitesimal volume centered around any point $\xv$ within the scatterer. With this replacement, the dependence on $\xv$ of each integrand becomes approximately constant (exactly constant in the zero-volume limit), and the integral simplifies to just multiplication by the volume $V_{\xv}$, which appears in every term and can be cancelled, leaving:
\begin{align}
    \int_V \Pv^*(\xv) \GG_0(\xv,\xv') \Pv(\xv') \dxv' + \Pv^*(\xv) \xi(\xv) \Pv(\xv) = -\Pv^*(\xv) \Einc(\xv).
    \label{eq:locoptthmxv}
\end{align}
More rigorous justifications are given in Refs.~\cite{Kuang2020,Molesky2020}, and can proceed either from the volume-integral formulation or, with equal validity, by converting the volume integrals around $V_{\xv}$ into surface integrals (via the divergence theorem), in which case \eqref{locoptthmxv} is interpreted simply as flux conservation through the surface of $V_{\xv}$. To convert \eqref{locoptthmxv} to the more compact vector notation, we denote new matrices $\DD_i$ as diagonal matrices of all zeros except a single 1 at diagonal entry $i$, in which case \eqref{locoptthmxv} can be written
\begin{align}
    \pv^\dagger \DD_i \left(\GG_0 + \xi\right) \pv = -\einc^\dagger \DD_i \xv,
    \label{eq:locoptthm}
\end{align}
which must hold for all spatial locations index by $i$. \Eqref{locoptthm} offers an infinite set of local conservation laws\index{local conservation laws} that must be satisfied for any (linear) scattering body. Moreover, just as for the global conservation laws\index{global conservation laws}, \eqref{locoptthm} is domain oblivious\index{domain oblivious}. Hence if the constraints of \eqref{locoptthm} lead to a bound, then that bound will apply to all sub-domains (or ``patterns'') contained therein. 

There is a systematic procedure that one can follow for identifying fundamental limits using the constraints of \eqref{locoptthm}. If one discards the Maxwell differential (or integral) equations, and only imposes the constraints of \eqref{locoptthm}, the resulting optimization problem has the form of a \textbf{quadratically constrained quadratic program}\index{quadratically constrained quadratic program (QCQP)}, or \textbf{QCQP}. QCQPs arise across many areas of science and engineering~\cite{Boyd1994,Goemans1995,Luo2010,Sojoudi2012,Candes2013,Horstmeyer2015}, and there are many mathematical approaches for solving them. One in particular is useful for identifying bounds: one can relax a QCQP to a \textbf{semidefinite program (SDP)}\index{semidefinite program (SDP)} in a higher-dimensional space~\cite{Laurent2005,Luo2010}, which can be solved for its global optimum by standard algorithms in polynomial time~\cite{Vandenberghe1996,Boyd2004}. The solution of the SDP\index{semidefinite program (SDP)} is guaranteed to be a bound, or fundamental limit, on the solution of the problem of interest. (The semidefinite program can also be regarded as the ``dual''\index{dual problem}~\cite{Boyd2004} of the dual\index{dual problem} of the QCQP\index{quadratically constrained quadratic program (QCQP)}~\cite{park_general_2017}, which is another way to see that it leads to bounds.)

Thus local conservation laws\index{local conservation laws} lead to a systematic procedure for identifying bounds, or fundamental limits, to electromagnetic quantities of interest. One replaces the governing Maxwell equations with the domain-oblivious\index{domain oblivious} conservation-law constraints of \eqref{locoptthm}, forms a semidefinite program\index{semidefinite program (SDP)} from the objective and constraints, and solves the SDP\index{semidefinite program (SDP)} to find a bound. To avoid the computational complexity of using all of the constraints, one can iteratively select only the ``maximally violated'' constraints, for rapid convergence to the bound of interest~\cite{Kuang2020}. A mathematically oriented review of bounds related to \eqref{locoptthm} is given in \citeasnoun{Angeris2021}. Extensions of various types are given in \citeasnoun{Shim2021subm} (multi-functionality), \citeasnoun{Zhang2021} (quantum optimal control), \citeasnoun{Angeris2022} (efficiency metrics), and \citeasnoun{Angeris2022b} (other physical equations).

\subsection{Sum rules}
\label{sec:sumrules}
Whereas the three previous sections primarily emphasized fundamental limits across spatial degrees of freedom, at a single frequency, \textbf{sum rules}\index{sum rules} center around \emph{spectral} degrees of freedom and constraints related to bandwidth. Sum rules are a prime example of applied complex analysis. Most often they are taught and discussed in the context of material susceptibilities, so we will start there, before focusing on our key interest, scattering problems. In the Appendix \secref{appCA} we provide a short review of key results from complex analysis, and the intuition behind their derivations, culminating in the Cauchy residue theorem\index{Cauchy residue theorem} that is used for all sum rules\index{sum rules}. Cauchy's residue theorem\index{Cauchy residue theorem}, for our purposes, can be distilled to the following statement. Consider a function $f(z)$ that is \textbf{analytic} (has no poles) in some domain $D$ in the complex $z$ plane. (Below, the analytic variable $z$ will be the frequency $\omega$.) Then the function $f(z) / (z-z_0)$ has a simple pole at $z_0$, for $z_0$ in $D$, and any integral of this function along a closed contour in $D$ containing $z_0$ simplifies to the value of the function at the pole:
\begin{align}
    \oint_{\gamma} \frac{f(z)}{z - z_0} = 2\pi i f(z_0),
    \label{eq:Cauchy}
\end{align}
where $f(z_0)$ is the ``residue'' of the function $f(z) / (z-z_0)$. Now let us put Cauchy's residue theorem\index{Cauchy residue theorem} to use.

Consider a material susceptibility $\chi$ that relates an electric field $\Ev$ to an induced polarization field $\Pv$. Typically we might directly consider the frequency-domain relationship of these variabes,
\begin{align}
    \Pv(\omega) = \chi(\omega) \Ev(\omega),
\end{align}
where we are suppressing spatial dependencies in these expressions for simplicity. (All of the position dependencies are straightforward.) This multiplicative frequency-domain relation arises from a \emph{convolutional} time-domain relationship: the polarization field at a given field is related to the electric field at all other times convolved with the susceptibility function (as a function of time):
\begin{align}
    \Pv(t) = \int \chi(t-t') \Ev(t') \,{\rm d}t'.
\end{align}
(We do not use different variables for the time- and frequency-domain definitions; the domain should be clear in each context.) \textbf{Causality}\index{causality} is the formal specification that \emph{cause precedes effect}. Material susceptibilities are causal: the polarization field cannot arise before the electric field has arrived, which means that for some origin of time, the susceptibility function is identically zero at all preceding times:
\begin{align}
    \chi(t-t') = 0 \qquad \textrm{ for } t < t'.
\end{align}
In the usual Fourier-transform relation between the time- and frequency-domain susceptibility functions, then, one can set the lower limit of the time-domain integral to be 0:
\begin{align}
    \chi(\omega) = \frac{1}{2\pi} \int_{-\infty}^{\infty} \chi(t) e^{i\omega t} \,{\rm d}t = \frac{1}{2\pi} \int_0^{\infty} \chi(t) e^{i\omega t} \,{\rm d}t.
    \label{eq:chiCaus}
\end{align}
Setting the lower limit of the integral to 0 has an important ramification. Let us assume the susceptibility takes a finite value for all real frequencies. (Metals are an exception, with divergent susceptibilities at zero frequencies, but known modifications to the rules below can be developed to account for this singularity~\cite{King1976,Lucarini2005}.) This implies that the integral of \eqref{chiCaus} converges to the correct finite value at each frequency. Now let us consider a \textbf{complex-valued} frequency $\omega = \omega_0 + i\Delta \omega$. If we insert this frequency into \eqref{chiCaus}, we find:
\begin{align}
    \chi(\omega_0+i\Delta\omega) = \frac{1}{2\pi} \int_0^{\infty} \chi(t) e^{i\omega_0 t}e^{-\Delta \omega t} \,{\rm d}t,
    \label{eq:chiwc}
\end{align}
which is equivalent to the integral of \eqref{chiCaus}, except now there is the additional exponential decay term $e^{-\Delta \omega t}$ in the integrand. This exponential decay term can only aid in convergence, and under appropriate technical assumptions (e.g. Titchmarsh's theorem~\cite{Nussenzveig1972}), one can prove the intuitive idea that \eqref{chiwc} cannot diverge for any $\Delta \omega$. This implies that the material susceptibility $\chi(\omega)$ is analytic in the upper-half of the complex-frequency plane. (Conversely, frequencies in the low half would have the exponentially diverging term $e^{\Delta \omega t}$ in their integrands, which would lead to divergences at certain frequencies, which is where the system \emph{resonances} are located.) Hence we can use the Cauchy integral theorem\index{Cauchy residue theorem} of \eqref{Cauchy} with $\chi(\omega)$ as the analytic function in the numerator of the integrand. The typical usage of the integral theorem is to select a pole on the real axis (or, technically, in the limit of approaching the real axis from above), and to use a contour $C$ that follows the real line, includes a semi-circular deformation around $\omega'$, and then closes along a semicircle approaching infinity in the upper-half plane. This contour actually does not enclose \emph{any} poles, instead ``side-stepping'' the real-axis pole, at a frequency we denote by $\omega$. Hence we have
\begin{align}
    \oint_C \frac{\chi(\omega')}{\omega' - \omega} \dw' = 0.
\end{align}
The integral over $C$ can be broken into three components: the principal-valued integral along the real axis from negative infinity to infinity (skipping $\omega'$), the semicircular arc going into the upper-half plane, and the semicircular arc rotating clockwise around $\omega$. The second of these terms is zero (for sufficient decay of $\chi(\omega)$), while the third term is simply $-i\pi \chi(\omega)$ (half of the typical Cauchy residue term since it is half of a circle, with a negative sign for the clockwise rotation). Equating the negative of the third term to the first, we have:
\begin{align}
    i\pi \chi(\omega) = \int_{-\infty}^{\infty} \frac{\chi(\omega')}{\omega' - \omega} \dw'.
\end{align}
We can take the imaginary part of both sides, and use the symmetry of $\chi$ around the origin, $\chi(-\omega) = \chi^*(\omega)$, to arrive at one of the \textbf{Kramers--Kronig} (KK)\index{Kramers--Kronig (KK) relations} relations for a material susceptibility:
\begin{align}
    \Re \chi(\omega) = \frac{2}{\pi}\int_0^{\infty} \frac{\omega' \Im \chi(\omega')}{(\omega')^2 - \omega^2} \dw'.
    \label{eq:chiKK}
\end{align}
The counterpart KK relation\index{Kramers--Kronig (KK) relations} relates the imaginary part of $\chi(\omega)$ to an integral involving the real part. These KK relations\index{Kramers--Kronig (KK) relations} are the foundations of sum rules\index{sum rules}. There are two special pole frequencies $\omega$ at which we may have additional information about the material response: infinity frequency and zero frequency (statics). In the limit of infinitely large frequencies, all materials become transparent, with a susceptibility that must scale as 
\begin{align}
    \chi(\omega) \rightarrow -\frac{\omega_p^2}{\omega^2} \qquad \textrm{ as } \omega \rightarrow \infty,
\end{align}
where $\omega_p$ is a constant proportional to the total electron density of the material~\cite{King1976,Lucarini2005}. Inserting this asymptotic limit into the KK relation\index{Kramers--Kronig (KK) relations} of \eqref{chiKK}, we find our first example of a sum rule\index{sum rules}:
\begin{align}
    \int_0^\infty \omega \Im \chi(\omega) \dw = \frac{\pi\omega_p^2}{2}.
    \label{eq:chiSumInf}
\end{align}
\Eqref{chiSumInf} is known as either the \textbf{TRK sum rule}\index{sum rules} or the \textbf{$f$ sum rule}\index{sum rules}~\cite{King1976,Lucarini2005}. It relates the weighted integral of the imaginary part of the susceptibility to simple constants multiplied by the electron density of the material of interest. The quantity $\omega \Im \chi(\omega)$ is proportional to the \emph{oscillator strengths} in single-electron susceptibility models~\cite{Kaxiras2019}. Alternatively, in the low-frequency limit, one may know the static refractive index $n_0$ of a given material; inserting $\omega=0$ in the KK relation\index{Kramers--Kronig (KK) relations} of \eqref{chiKK} gives the low-frequency sum rule\index{sum rules}:
\begin{align}
    \int_0^\infty \frac{\Im \chi(\omega)}{\omega} \dw = \frac{\pi}{2} \left(n_0^2 - 1\right).
    \label{eq:chiSum0}
\end{align}
The two sum rules\index{sum rules} of \eqreftwo{chiSumInf}{chiSum0} are well-known material sum rules\index{sum rules} that are useful for spectroscopy~\cite{King1976,Lucarini2005} as well as for bounds on material properties~\cite{Skaar2006,Gustafsson2010,Shim2021}. We have repeated their well-known derivations to familiarize the reader with the machinery of KK relations\index{Kramers--Kronig (KK) relations} and sum rules\index{sum rules}, which we apply next to scattering properties.

Just as the origin for material sum rules\index{sum rules} was recognition of material susceptibility as a causal (linear) response function, for scattering sum rules\index{sum rules} we want to start by recognizing that the electromagnetic field $\Ev$ generated by a source (presumably current) is also a causal linear response function: $\Ev$ cannot be nonzero before the current $\Jv$ is nonzero. Hence the electric field at all times before an origin must be zero, which again leads to analyticity in the upper-half of the complex-frequency plane. Yet we do not want KK relations\index{Kramers--Kronig (KK) relations} for the electric field at specific points in space; we want KK relations\index{Kramers--Kronig (KK) relations} (and sum rules\index{sum rules}) for relevant power quantities. Typical expressions of interest might be the field intensity, $|\Ev(\xv,\omega)|^2$, or the Poynting flux $(1/2)\Re\left[\Ev^*(\xv,\omega) \times \Hv(\xv,\omega)\right]$, at a point $\xv$, but \emph{neither} of these quantities is analytic in the upper-half plane. The problematic term in each case is $\Ev^*(\omega)$. Analyticity is not preserved under complex conjugation, and indeed by symmetry we know that $\Ev^*(\omega) = \Ev(-\omega)$ on the real line; if we try to continue $\omega$ into the upper-half plane, the $-\omega$ argument moves into the lower-half plane, where the resonances reside. Hence $\Ev^*(\omega)$ can have poles, and the corresponding power terms do not have simple KK relations\index{Kramers--Kronig (KK) relations} or sum rules\index{sum rules}.

We are rescued, again, by the optical theorem\index{optical theorem}. Whereas absorbed and scattered powers always involve conjugated total fields, \emph{extinction}, by virtue of the optical theorem\index{optical theorem}, takes a different form (\eqref{Pext}), which is proportional to the overlap integral of the conjugate of the incident field with the induced polarization field, $\int_V \Einc^* \cdot \Pv$. Many common incident fields, such as plane waves of the form $e^{i\omega x / c}$, are analytic \emph{everywhere} in the complex plane, and their conjugates \emph{can} be analytically continued. The polarization field is the product of the analytic material susceptibility with the analytic electric field, and thus is itself analytic. Hence extinction expressions contain a term that will obey KK relations\index{Kramers--Kronig (KK) relations} and sum rules\index{sum rules}, which we denote $s(\omega)$:
\begin{align}
    \Pext(\omega) = \frac{\omega}{2} \Im \underbrace{\int_V \Einc^*(\xv,\omega) \cdot \Pv(\xv,\omega) \dxv}_{s(\omega)}.
    \label{eq:Ps}
\end{align}
By the arguments laid out above, the quantity $s(\omega)$ is analytic in the upper-half plane. It satisfies the other required assumptions as well (e.g. sufficient decay at infinity) for incident fields such as plane waves; we can immediately write a KK relation\index{Kramers--Kronig (KK) relations} for it:
\begin{align}
    \Re s(\omega) = \frac{2}{\pi} \int_0^\infty \frac{\omega' \Im s(\omega')}{(\omega')^2 - \omega^2} \dw'.
    \label{eq:sKK}
\end{align}
Notice that the term in the numerator of the integrand is exactly proportional to extinction; hence sum rules\index{sum rules} for the imaginary part of $s(\omega)$ (by analogy with the sum rules\index{sum rules} for $\Im \chi$) will necessarily be sum rules\index{sum rules} for extinction. Again paralleling the susceptibility analysis, we can take the limit as $\omega \rightarrow \infty$, in which case
\begin{align}
    s(\omega) &= \int \Einc^*(\xv,\omega) \cdot \Pv(\xv,\omega) \dxv \nonumber \\
              &\rightarrow -\frac{\omega_p^2}{\omega^2} \int_V \left| \Einc(\xv,\omega) \right|^2 \dxv \nonumber \\
              &= -\frac{\omega_p^2}{\omega^2} \left| \Ev_0 \right|^2 V,
\end{align}
where $\Ev_0$ is the (constant) vector amplitude of the plane wave, and $V$ is the volume of the scatterer. Evaluating the KK relation\index{Kramers--Kronig (KK) relations} for $s(\omega)$, \eqref{sKK}, in the high-frequency limit gives a sum rule\index{sum rules} for the imaginary part of $s(\omega)$:
\begin{align}
    \int_0^\infty \omega \Im s(\omega) \dw = \frac{\pi \omega_p^2}{2} \left|\Ev_0\right|^2 V,
\end{align}
which in turn implies a sum rule\index{sum rules} for extinction (via \eqref{Ps}):
\begin{align}
    \int_0^\infty \Pext(\omega) \dw = \frac{\pi \omega_p^2}{4} \left|\Ev_0\right|^2 V.
    \label{eq:extSumInf}
\end{align}
\Eqref{extSumInf} dictates that the total integrated extinction of any scattering body is fixed by the amplitude of the incident plane wave and the total number of electrons in the scatterer (from the product of $\omega_p^2$ with $V$), and is otherwise independent of the shape, resonance profile, and any other characteristics of the scattering body.

Just as for a material susceptibility, one can also derive a sum rule\index{sum rules} for $\Pext$ by setting $\omega=0$ in the KK relation\index{Kramers--Kronig (KK) relations} for $s(\omega)$, \eqref{sKK}. The key low-frequency information we can utilize is that the net induced dipole moment of the scatterer is related to the incident field via a polarizability tensor $\alphat$. Following a few algebraic steps~\cite{Sohl2007} paralleling the low-frequency material sum rule\index{sum rules}, one similarly finds a sum rule\index{sum rules} for the integrral of $\Pext(\omega) / \omega^2$. The term $(1/\omega^2) \dw$ is exactly proportional to ${\rm d}\lambda$, where $\lambda = 2\pi c/\omega$ is the wavelength, so this sum rule\index{sum rules} is often written as a sum rule\index{sum rules} over wavelength:
\begin{align}
    \int_0^\infty \Pext(\omega) \,{\rm d}\lambda = \pi^2 \Ev_0 \cdot \alphat \Ev_0.
    \label{eq:extSum0}
\end{align}
There is an additional magnetic polarizability term in materials with a nonzero magnetostatic response~\cite{Sohl2007}. Interestingly, \eqref{extSum0} has different dependencies than \eqref{extSumInf}: the polarizability has a weak dependence on material, but a strong dependence on shape. The low-frequency sum rule\index{sum rules} implies that scattering bodies with the same size and shape, but made of different materials, can have nearly identical wavelength-integrated extinctions. Moreover, electrostatic polarizabilities obey ``domain monotonicity'' bounds that dictate that the quantity $\Ev_0 \cdot \alphat \Ev_0$ must increase as the scatterer domain increases in size, such that one can bound integrated extinction via high-symmetry enclosures for which the right-hand side of \eqref{extSum0} often takes a simplified analytical form. Taken together, the high- and low-frequency sum rules\index{sum rules} of \eqreftwo{extSumInf}{extSum0} comprise strong constraints on the possible scattering lineshapes of arbitrary scatterers.

\eqreftwo{extSumInf}{extSum0} are classical sum rules\index{sum rules} with a long history. The high-frequency sum rule\index{sum rules}, \eqref{extSumInf}, was known at least as early as 1963~\cite{Gordon1963}, when the connection to material-susceptibility sum rules\index{sum rules} was first made. A specialized version of the low-frequency sum rule\index{sum rules}, \eqref{extSum0}, was first proposed by Purcell in 1969~\cite{Purcell1969}, in order to bound the minimum volume occupied by interstellar dust. It was generalized to arbitrary scattering bodies in \citeasnoun{Sohl2007}, where the monotonicity bounds (originally developed by Jones~\cite{Jones1985}) were connected to the low-frequency sum rules\index{sum rules}. For many years, it seemed that plane-wave extinction might be the \emph{only} scattering quantity for which sum rules\index{sum rules} can be derived. In recent years, however, it has been recognized that near-field\index{near field} local density of states has a similar form---it is the real or imaginary part of an amplitude, instead of the squared magnitude of an amplitude---for which sum rules\index{sum rules} can also be derived. We describe this sum rule\index{sum rules} and its implications in \secref{allfreqsumrules}.

\section{Fundamental limits in the near field}
\label{sec:NearFieldLimits}
We have set the stage: we have introduced near-field\index{near field} optics, defined many of the response functions of interest, and described tools formulated for electromagnetic-response bounds. In this section we describe how these ingredients come together for bounds and fundamental limits to near-field\index{near field} response. We identify different bounds---and the different techniques required to derive them---based on the frequency range of interest: a single frequency (\secref{SFBounds}), all frequencies (\secref{allfreqsumrules}), and finite, nonzero bandwidths (\secref{finiteband}). We leave bounds for mode volume\index{mode volume}, which seemingly requires very different techniques, to the final section of the chapter (\secref{modevbounds}).

\subsection{Single-frequency bounds}
\label{sec:SFBounds}
In \secref{BoundApproaches}, we described two techniques that can be used to identify single-frequency bounds to any linear-electromagnetic response function of interest: conservation laws and channel decompositions. In this subsection we summarize how one can adapt, specialize, and/or combine those approaches in the near field\index{near field}, for spontaneous-emission\index{spontaneous emission} and CDOS\index{cross density of states (CDOS)} engineering, Smith--Purcell radiation enhancements\index{Smith--Purcell radiation}, and spectral NFRHT\index{near-field radiative heat transfer (NFRHT)} response. 

\subsubsection{Spontaneous emission}
\label{sec:SEbounds}
The canonical near-field\index{near field} quantity is LDOS\index{local density of states (LDOS)}, which as discussed in \secref{LDOS} is proportional to the spontaneous emission\index{spontaneous emission} rate of an electric dipole at a given location. In a closed system, the LDOS is a sum of delta functions over the modes of the system, in which case the LDOS diverges at the modal frequencies. In an open system, however, the modal intuition no longer applies, leading to the more general Green's-function expression of \eqref{LDOSe}. This scattering quantity lends itself well to the conservation-law-based scattering-response bounds described in \secref{local}. 

We can repeat here the Green's function expression for LDOS, which we will denote in this section by $\rho(\xv,\omega)$:
\begin{align}
    \rho(\xv,\omega) = \frac{1}{\pi\omega} \Tr \Im \GG(\xv,\xv,\omega).
\end{align}
The trace of the Green's function can be computed with a summation over three orthogonal unit vectors $\vect{s}_j$, for $j=1,2,3$, in which case the trace can be interpreted as the incoherent summation of the fields from three dipoles with amplitudes $\varepsilon_0 \vect{s}_j$. There is an initial impediment to applying the conservation-law framework to this expression: it is not written explicitly as a function of the polarization fields, whose constraints are critical to meaningful bounds. This impediment is easily hurdled: one can decompose the Green's function into its incident and scattered components. The scattered fields are the convolutions of the free-space Green's-function matrix $\GG_0$ from the scattering domain to the dipole point; by reciprocity, the overlap of $\vect{s}_j$ with $\GG_0$ is the field incident upon the scattering body $V$. By this line of reasoning, for a scalar isotropic medium (the general bianisotropic case is derived in \citeasnoun{Miller2016}), one can rewrite LDOS\index{local density of states (LDOS)} as
\begin{align}
    \rho(\xv,\omega) = \rho_0(\omega) + \frac{1}{\pi\omega} \Im \sum_j \int_V \Ev_{\textrm{inc},\vect{s}_j} \cdot \Pv_{s_j} \,{\rm d}V,
\end{align}
where $\rho_0(\omega)$ is the free-space LDOS (which is position-independent, and given below \eqref{LDOSe}), and the $\vect{s}_j$ subscript encodes the three dipole orientations. Using the same discretized vector/matrix notation as we initiated with \eqref{optthm}, this expression can equivalently be written
\begin{align}
    \rho(\xv,\omega) = \rho_0(\omega) + \frac{1}{\pi\omega} \Im \sum_j \ev_{\textrm{inc},\vect{s}_j}^T \pv_{s_j}.
    \label{eq:LDOSscat}
\end{align}
Now we see that LDOS is a linear function of the polarization fields induced in the scattering body. We want to know the largest possible value of LDOS, of \eqref{LDOSscat}, subject to Maxwell's equations, but of course the latter constraint contains all of the complexity of the design problem. Instead, we drop the Maxwell-equation constraint, and impose only one of the conservation laws of \secref{BoundApproaches}. To start, we can impose the conservation law that absorbed power be smaller than extinguished power, of \eqref{optthmabs}, which leads to the optimization problem:
\begin{equation}
    \begin{aligned}
        & \underset{\pv_{\vect{s}_j}}{\text{max.}}
        & & \frac{1}{\pi\omega} \Im \sum_j \ev_{\textrm{inc},\vect{s}_j}^T \pv_{\vect{s}_j} \\
        & \text{s.t.}
        & & \left(\Im \xi\right) \pv_{\vect{s}_j}^\dagger \pv_{\vect{s}_j} \leq \Im \left(\ev_{\textrm{inc},\vect{s}_j}^\dagger \pv_{\vect{s}_j} \right).
    \end{aligned}
    \label{eq:opt_prob}
\end{equation}
Treating each dipole orientation $\vect{s}_j$ independently, one can find from a Lagrangian analysis that the optimal $\pv_{\vect{s}_j}$ comprises a linear combination of $\ev_{\textrm{inc},\vect{s}_j}$ and $\ev^*_{\textrm{inc},\vect{s}_j}$; in the near field\index{near field}, where the incident field and its conjugate are nearly identical, and the LDOS\index{local density of states (LDOS)} is dominated by its scattered-field contribution,  we ultimately find the following bound~\cite{Miller2016}:
\begin{align}
    \rho(\xv,\omega) \leq \frac{1}{\pi\omega} \frac{|\chi(\omega)|^2}{\Im \chi(\omega)} \sum_{\vect{s}_j} \left\| \ev_{\textrm{inc},\vect{s}_j} \right\|^2 = \frac{1}{\pi\omega} \frac{|\chi(\omega)|^2}{\Im \chi(\omega)} \sum_{\vect{s}_j} \int_V \left| \Ev_{\textrm{inc},\vect{s}_j} \right|^2 \dxv.
    \label{eq:rhobound0}
\end{align}
Normalizing by the free-space electric LDOS $\rho_0(\omega)$, and performing the integral over an enclosing half-space (and keeping ony the term that decreases most rapidly with separation distance $d$), one finds~\cite{Miller2016}:
\begin{align}
    \frac{\rho(\xv,\omega)}{\rho_0(\omega)} \leq \frac{1}{8(kd)^3} \frac{|\chi(\omega)|^2}{\Im \chi(\omega)},
    \label{eq:rhobound}
\end{align}
where $k=\omega/c$ is the free-space wavenumber. \Eqref{rhobound} represents our first near-field\index{near field} bound. This bound only depends on two parameters of the system: the separation distance $d$, relative to the wavenumber, and the \textbf{material enhancement factor}\index{material enhancement factor},
\begin{align}
    \frac{|\chi(\omega)|^2}{\Im \chi}.
\end{align}
The material enhancement factor\index{material enhancement factor} encodes a key tradeoff: a large susceptibility magnitude implies large possible polarization currents, while a large imaginary part of the susceptibility implies losses that necessarily restrict resonant enhancement. In Drude metals with $\chi = -\omega_p^2 / (\omega^2 + i\gamma\omega)$, the material enhancement factor\index{material enhancement factor} is given by $\omega_p^2 / \gamma\omega$, showing that the largest possible single-frequency response is achievable in materials with large electron densities and small losses. The material enhancement factor\index{material enhancement factor} is described in further detail in Refs.~\cite{Miller2016,Shim2020b}.

The second key parameter is the distance $d$; the factor $1/d^3$ encodes the dramatic enhancements that are possible in the near field\index{near field}. These enhancements are typically achieved with plasmonic\index{plasmons} modes, and the factor $1/d^3$ arises from the most rapidly decaying component of the free-space Green's function, $\sim 1/r^3$; squaring this term and integrating over a three-dimensional volume leads to the inverse-cubic dependence. The last point also suggests an important caveat: systems with a different dimensionality must have different scaling laws as a function separation distance. Designing for 2D materials, for example, leads to integrals over 2D (or very thin) domains, leading to a $1/d^4$ near-field\index{near field} enhancement factor. There are also more slowly increasing terms that arise from the mid-field and far-field contributions to the free-space Green's function.

Finally, it should be noted that certain constraints of interest can be seamlessly integrated into the optimization problem of \eqref{opt_prob}. Of particular importance in plasmonics\index{plasmons} applications is \textbf{radiative efficiency}\index{radiative efficiency}. When one finds a bound on extinction or LDOS\index{local density of states (LDOS)}, the bound may suggest very large enhancements, but all of that enhancement could be going into material absorption rather than far-field radiation or scattering. Suppose a given application requires a certain radiative efficiency\index{radiative efficiency}, such as some fraction $\eta$ of the total emission going into the far field. This can be written mathematically as the constraint that absorption be smaller than $(1-\eta)$ multiplied by the extinction, or $\Pabs \leq (1-\eta) \Pext$. Absorption is quadratic in the polarization field, while extinction is linear in the polarization field, such that this expression represents an additional constraint that can be seamlessly incorporated into \eqref{opt_prob}. Often the bound of interest, with this constraint, is analytically solvable. \citeasnoun{Yang2017} identifies precisely such bounds on high-radiative-efficiency\index{radiative efficiency} plasmonics\index{plasmons}, prescribing a tradeoff between large response and radiative efficiency\index{radiative efficiency}. In \citeasnoun{Yang2017} it is not only shown that high-radiative-efficiency\index{radiative efficiency} bounds can be derived; it is also shown that hybrid dielectric-metal designs can approach the bounds, \emph{and} that they surpass the same fundamental limits evaluated for metal-only structures. This example showcases the power of using bounds to understand the broader landscape of a photonics application area of interest.

\subsubsection{CDOS}
Bounds to CDOS\index{cross density of states (CDOS)} can be found along very similar lines to the LDOS bounds of above. We can define the trace of the CDOS via \eqref{CDOS}, taking
\begin{align}
    \rho(\xv_1,\xv_2,\omega) = \frac{1}{\pi\omega} \Tr \Im \GG(\xv_1,\xv_2,\omega).
\end{align}
Then, we can separate out a scattered contribution coming from the polarization fields induced in the scatterer, just as for LDOS, and when this term dominates (i.e. the geometry primarily mediates the CDOS), we have:
\begin{align}
    \rho(\xv_1,\xv_2,\omega) = \frac{1}{\pi\omega} \sum \ev_{\textrm{inc},\vect{s}_j,\xv_1}^T \pv_{\vect{s}_j,\xv_2},
\end{align}
where the position subscripts on $\ev_{\rm inc}$ and $\pv$ denote the source positions of the $\vect{s}_j$-polarized dipoles. Hence in CDOS the field incident from one position is overlapped with the polarization field induced by a source from a second position. The bound for CDOS will be identical to that of \eqref{rhobound0}, but with $\|\ev_{\textrm{inc},\vect{s}_j}\|^2$ replaced by $\|\ev_{\textrm{inc},\vect{s}_j,\xv_1}\| \|\ev_{\textrm{inc},\vect{s}_j,\xv_2}\|$. Finally, normalizing by free-space LDOS and dropping all except the most rapidly varying terms as a function of separation distances $d_1$, $d_2$, one arrives at the bound~\cite{Shim2019}
\begin{align}
    \frac{\rho(\xv_1,\xv_2,\omega)}{\rho_0(\omega)} \leq \frac{1}{4k^3\sqrt{d_1^3 d_2^3}} \frac{|\chi(\omega)|^2}{\Im \chi(\omega)}.
    \label{eq:rho12bound}
\end{align}
The discussion of the terms that appeared in the LDOS bound of \eqref{rhobound} can be translated almost seamlessly here: the same material dependence shows up, corresponding to the same possibilities for plasmonic\index{plasmons} enhancement, and the same distance dependencies due to the same enhancements of the near fields\index{near field} of the two dipoles. There are likely two further enhancements that can be made to \eqref{rho12bound}. First, \eqref{rho12bound} is a factor of 2 larger than \eqref{rhobound}, when the former is evaluated in the limit as $\xv_1 \rightarrow \xv_2$. This is almost certainly because the bound of \eqref{rho12bound} in \citeasnoun{Shim2019} came from evaluating bounds for each diagonal element, simplifying, \emph{then} taking the trace. Taking the trace and then simplifying the bound would likely remove this factor of 2. Second, the bound of \eqref{rho12bound} does \emph{not} depend on the distance between the two dipoles, $d_{12}$. This may be physical in certain limits, e.g. when a plasmon\index{plasmons} can maintain its amplitude in propagating from one dipole to the other, but may not be physical when such propagation is not possible, and one would expect improved bounds to capture this. It is likely true that applying the many-conservation-law approach of \secref{local} would incorporate such effects. Nevertheless, \eqref{rho12bound} is a good starting point to understand the upper limits to engineering CDOS\index{cross density of states (CDOS)} in photonic environments.

\subsubsection{Smith--Purcell radiation}
Another exciting application area for the single-frequency bound approach is to Smith--Purcell radiation\index{Smith--Purcell radiation}, which is the radiation that occurs when a free electron passes near a structured material. A constant-velocity free electron produces only a near field\index{near field}, with no far-field component, but when the evanescent wave interacts with grating-like structures, the gratings can couple the near fields\index{near field} to propagating far fields, leading to a release of energy from the electron in the form of electromagnetic rediation. The natural question, then, is how large this energy release can be?

Mathematically, this question is identical to the question of the work done by a dipole (i.e., LDOS\index{local density of states (LDOS)}), except that the incident field is different in this case, and is given by \eqref{EincFE}. Maximizing the overlap of \emph{this} incident field with the induced polarization field, subject to the same constraint of \eqref{opt_prob}, leads to a bound on the Smith--Purcell emission\index{Smith--Purcell radiation} spectral probability given by~\cite{Yang2018}
\begin{align}
    \Gamma(\omega) \leq \frac{\alpha}{2\pi c} \frac{|\chi|^2}{\Im \chi} \frac{L\theta}{\beta} \left[ (\kappa_\rho d) K_0(\kappa_\rho d) K_1(\kappa_\rho d) \right],
    \label{eq:SPBound}
\end{align}
where $\Gamma = P / \hbar \omega$ for emission power $P$, $\alpha$ is the fine-structure constant, $\beta = v/c$ is the normalized electron velocity, $L$ and $\theta$ are the height and opening azimuthal angle of the cylindrical sector containing the patterned material, $\kappa_{\rho} = k/\beta\gamma$ is the wavenumber divided by $\beta$ and the Lorentz factor $\gamma$, $d$ is the distance of the beam from the surface, and $K_n$ is the modified Bessel function of the second kind. Although the exact expression is somewhat complex, we see that Smith--Purcell radiation\index{Smith--Purcell radiation} also directly benefits from the material enhancement factor\index{material enhancement factor} $|\chi|^2 / \Im \chi$. A seemingly surprising conclusion also emerged from \eqref{SPBound}: \emph{slow} electrons, at small enough separations, can lead to \emph{greater} radiation enhancements than \emph{fast} (i.e. high-energy) electrons. All constant-velocity electrons do not radiate when their speed is smaller than speed of light in the background medium, and emit only near fields\index{near field}. But high-speed electrons are closer to surpassing the Cherenkov threshold, and hence the fields they generate decay more slowly, out to larger distances. By constrast, low-speed electrons have very strong but very tightly confined near fields\index{near field}. But if one brings a patterned surface close enough, the strong very near fields\index{near field} of slow electrons have greater potential for radiation enhancements than the more moderate near fields\index{near field} of fast electrons.

\begin{figure}
    \includegraphics[width=1\textwidth]{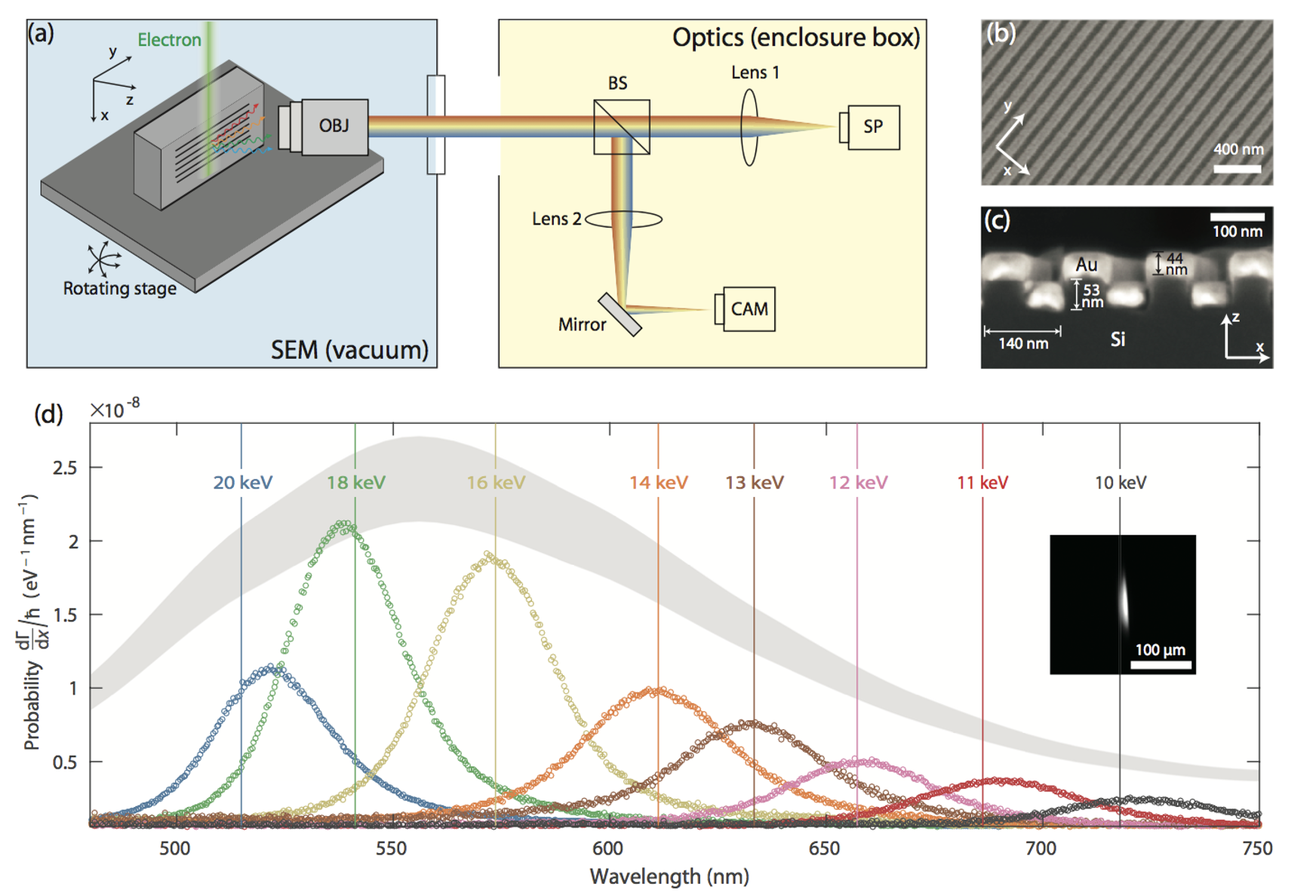}
    \centering
    \caption{The bounds of \eqref{SPBound} dictate upper limits to Smith--Purcell emission rates\index{Smith--Purcell radiation}. (a--d) The experiments of \citeasnoun{Yang2018} quantitatively confirm that designed metallic gratings can approach the fundamental performance limits. (Adapted from \citeasnoun{Yang2018}.)}
    \label{fig:SP}
\end{figure}

Some of the general trends, and absolute numerical values, of the bound of \eqref{SPBound} were validated theoretically and experimentally in \citeasnoun{Yang2018}. In particular, \figref{SP} shows an experimental setup for measuring the Smith--Purcell radiation\index{Smith--Purcell radiation} for electron beams with varying energies, as well as designed gold-on-silicon gratings whose parameters were optimized for maximum response. The key result is shown in panel (d), where the grey region indicates the fundamental bounds, as a function of photon wavelength, with some width to account for experimental undertainties. The colored data points are quantitatively measured probabilities (with no fitting parameters), showing that both the quantitative values of the bounds are nearly approachable, and that the complex wavelength dependence (emerging from an interplay between the material enhancement factor\index{material enhancement factor} and the optical near fields\index{near field}) correctly captures the response of high-performance designs.

\subsubsection{Spectral NFRHT}
\label{sec:NFRHTBounds}
Near-field radiative heat transfer, NFRHT\index{near-field radiative heat transfer (NFRHT)}, introduced in \secref{NFRHT}, offers an extraordinary challenge for fundamental limits. It comprises rapidly decaying, large-area, broadband thermal sources for which little has been understood about upper bounds for quite some time. While we tackle the question of broadband enhancements in \secref{finiteband}, in this section we describe the recent progress in understanding maximum NFRHT\index{near-field radiative heat transfer (NFRHT)} at a single frequency. There are three key results that we can highlight: channel bounds for planar bodies~\cite{Pendry1999,Biehs2010}, material-loss bounds~\cite{Miller2015}, and an amalgamation of the two~\cite{Molesky2020c,Venkataram2020}. 

Channel bounds to NFRHT\index{near-field radiative heat transfer (NFRHT)} are described as ``Landauer bounds,'' due to their similarities with Landauer transport. For planar bodies with in-plane translational (and therefor rotational) symmetries, the in-plane wavenumber is a conserved quantity, and the energy flux from one body to another can be decomposed into propagating and evanescent plane-wave channels with no cross-channel scattering. One can decompose the fields emanating from the emitting body into normalized plane-wave modes, insert them into the fluctuation-averaged flux, i.e. the average of the integral $\frac{1}{2} \int_A \Ev \times \Hv^* \cdot \nhat$, for separating plane $A$ and normal vector $\nhat$. This results in an expression for the flux rate $\Phi(\omega)$, of \eqref{H2r1} and \eqref{HTC}, given by
\begin{align}
    \Phi(\omega) = \sum_{j=s,p} \frac{1}{2\pi} \int \frac{{\rm d}^2\boldsymbol{\kappa}}{4\pi^2} T_j^{12}(\omega,\kappa,d),
    \label{eq:NFRHTPW}
\end{align}
where $\boldsymbol{\kappa}$ is the in-plane wave propagation constant (and $\kappa$ its magnitude), $j$ is a polarization index, $k_0$ is the free-space wavenumber and the $T_i$ are ``transmission coefficients,'' which depend on the specific Fresnel reflection coefficients of the two interfaces~\cite{Biehs2010}. This expression has an elegant interpretation: NFRHT\index{near-field radiative heat transfer (NFRHT)} is the composition of plane-wave fluxes, each contributing with a weight $T_i$. Moreover, the coefficients $T_i$ are bounded above by 1, for both the propagating and evanescent waves~\cite{Pendry1999,Biehs2010,Ben-Abdallah2010}. Then, if there is a limit to the largest wavenumber across which a nonzero transmission can be achieved, one will have a bound on the maximum spectral RHT. 

Hence it is possible to identify a maximal rate of NFRHT\index{near-field radiative heat transfer (NFRHT)} is given by power transferred with ``Landauer'' transmission unity over all possible plane waves~\cite{Pendry1999,Ben-Abdallah2010}. While intuitive, however, this bound has two serious drawbacks. The first is that if one literally computes the integral of \eqref{NFRHTPW} over all possible waves, the result is infinite, as there are an infinite number of plane-wave channels. Of course one cannot reasonably expect to achieve unity transmission over channels with infinitely large in-plane wavenumbers (as they decay exponentially fast), implying there must be a maximal channel at which the sum should be terminated. But how to choose this value? One proposal, from \citeasnoun{Pendry1999}, was that the maximal accessible channel should be proportional to $1/a$, where $a$ is the lattice spacing of the material; the reasoning being that beyond this limit the use of a continuum model of the materials would not be valid. Another proposal, from \citeasnoun{Ben-Abdallah2010}, is that the maximal accessible channel wavenumber is given by $k_{\rm max} = 1/d$, where $d$ is the separation between the two bodies; the reasoning being that the exponential decay of the evanescent waves makes it difficult to achieve large transmission beyond $1/d$. Each of the resulting bounds (one from $k_{\rm max} = 1/a$, and the other from $k_{\rm max} = 1/d$), has shortcomings: the lattice-spacing-defined bound is extraordinarily high for any reasonable lattice constant, well beyond all other bounds discussed below. And the separation-defined-bound is in fact not a true bound: it can be superceded with reasonable material parameters~\cite{Miller2015}, which in fact do show non-trivial transmission beyond $1/d$. Hence the two known versions of the bound are either far too large, or surpassable. 

The second serious drawback of using \eqref{NFRHTPW} is that it only applies to planar bodies with translational symmetry in all directions. The use of conservation laws for bounds, discussed next, leads to bounds that can apply to planar bodies with any patterning, while also being tighter than the channel bounds resulting from \eqref{NFRHTPW}.

The first use of conservation laws for spectral NFRHT\index{near-field radiative heat transfer (NFRHT)} bounds appeared in \citeasnoun{Miller2015}. The mathematical procedure is sufficiently complex that we will not go through it in detail here, but the intuition can be explained. The idea is to use the global conservation law\index{global conservation laws} requiring $\Pabs \leq \Pext$ in the spectral NFRHT\index{near-field radiative heat transfer (NFRHT)} problem. The difficulty is that the sources are embedded \emph{within} one of the scattering bodies, which leads to divergences if one blindly applies the constraint $\Pabs \leq \Pext$. However, the radiative exchange of heat can be decomposed into two subsequent scattering problems, both of which have sources separated from scatterers. In the first step, the incident field is given by the field emanating from body 1 \emph{in the presence of body 1}, with only the second body serving as the scatterer. The absorption in this second body is bounded by the extinction by this second body, which leaves a bound in terms of the second material and the ``incident field'' emanating from body 1. Of course, we do not know exactly what this field is for any pattern. At this point, however, we can use reciprocity to rewrite the field emanating from body 1 in terms of fields emanating from the free space of body 2's domain, being absorbed by body 1. The constraint $\Pabs \leq \Pext$ can be applied to this scattering process again, ultimately yielding a single-frequency, flux-per-area $A$ bound given by~\cite{Miller2015}
\begin{align}
    \frac{\Phi(\omega)}{A} \leq \frac{1}{16\pi^2 d^2} \frac{|\chi_1|^2}{\Im \chi_1} \frac{|\chi_2|^2}{\Im \chi_2},
    \label{eq:NFRHTSF}
\end{align}
where $d$ is the separation distance between the two bodies, and $\chi_{1}$ and $\chi_2$ are their optical susceptibilities, respectively.  This bound includes two key dependencies: the material enhancement factor\index{material enhancement factor} $|\chi|^2 / \Im \chi$, and a $1/d^2$ dependence arising from the rapidly decaying near fields\index{near field} in the electromagnetic Green's function. The bound of \eqref{NFRHTSF} is promising, as it suggests significant possible enhancements of spectral NFRHT\index{near-field radiative heat transfer (NFRHT)}, and it is plausible: the actual NFRHT of two planar bodies with equal susceptibilities, on resonance, is given by $\Phi(\omega)/A = 1/(4\pi^2 d^2) \ln\left[|\chi|^4/(4(\Im \chi)^2)\right]$, with nearly identical dependencies as \eqref{NFRHTSF}, except for the logarithmic dependence on the material enhancement. Can this be overcome, with instead linear enhancements in $|\chi|^2 / \Im \chi$? For some materials, the answer is ``yes,'' as shown with computational inverse design in \citeasnoun{Jin2019}. More generally, however, such linear enhancements are not generic, and one can further tighten the bound of \eqref{NFRHTSF}.

\begin{figure}[tb]
    \centering\includegraphics[width=0.99\linewidth]{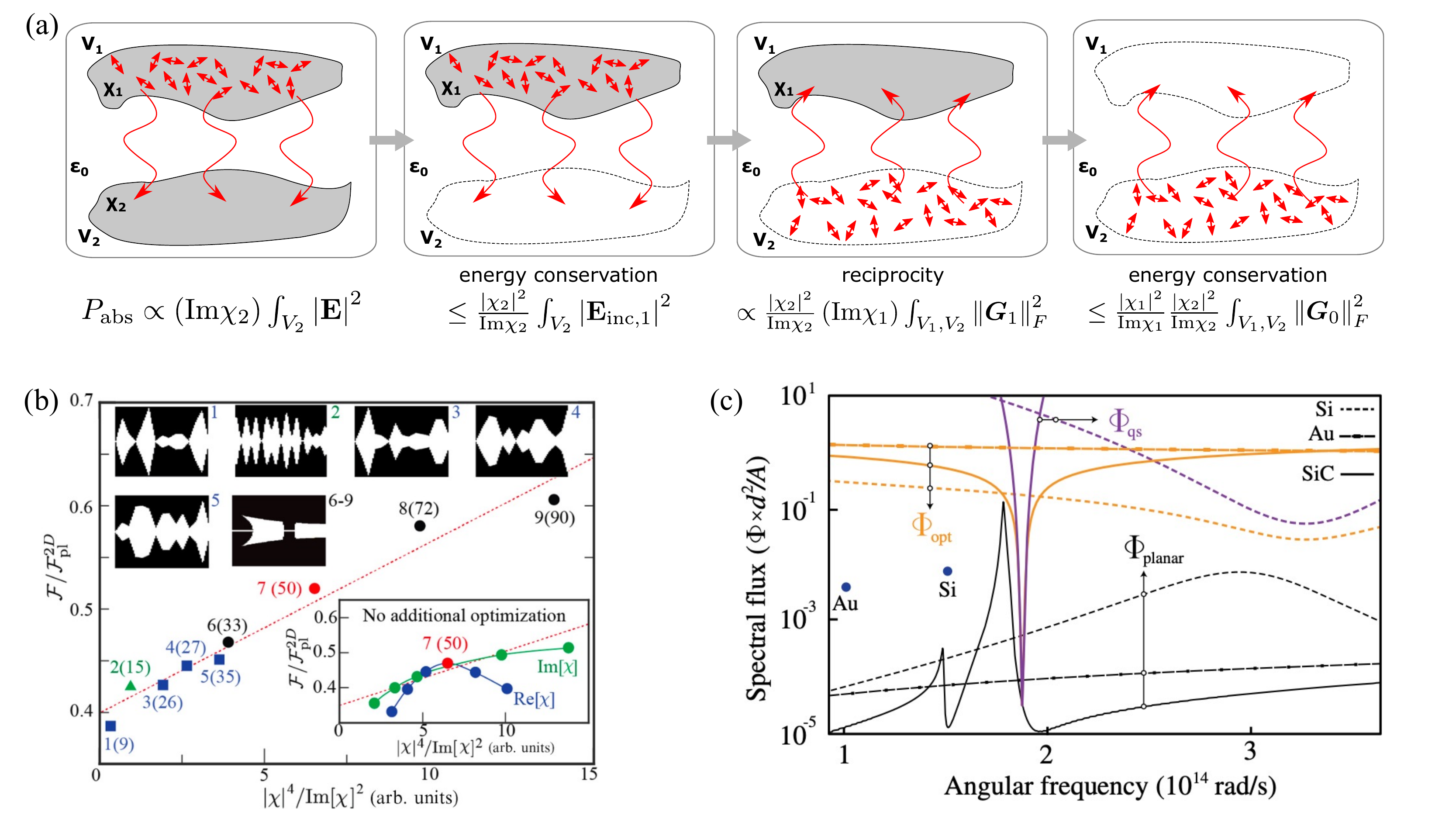}
    \caption{A collection of bounds on single-frequency near-field\index{near field} radiative heat transfer. (a) The approach of \citeasnoun{Miller2015} using material loss as the only constraint, exploiting reciprocity to bound the response given that the sources are embedded \emph{within} one of the arbitrarily patterned scattering bodies. (Adapted from \citeasnoun{Miller2015}.) (b) Bounds and designs from \citeasnoun{Jin2019} showing the feasibility, in specific regimes, of achieving enhancements proportional to the square of the material enhancement factor\index{material enhancement factor} $|\chi|^2 / \Im \chi$. (Adapted from \citeasnoun{Jin2019}.) (c) Tightened bounds from Refs.~\cite{Molesky2020c,Venkataram2020}, precluding the possibility of extraordinary response at frequencies away from the surface-polariton\index{polaritons} frequency of a material of interest. (Adapted from \citeasnoun{Venkataram2020}.)}
    \label{fig:SpectralNFRHT}
\end{figure}

Refs.~\cite{Molesky2020c,Venkataram2020} showed that one can tighten the bound of \eqref{NFRHTSF} by combining the use of a global conservation law\index{global conservation laws} with that of a channel decomposition. If one decomposes the \emph{general} (not specific to translation-symmetric) scattering response into plane waves, and further imposes conservation laws for absorption and extinction (of the bodies in tandem as well as in isolation), then a long mathematical process leads to a tighter bound. If we define $\GG_{0,AB}$ to be the free-space Green's function matrix for sources in body $A$ to measurement points in body $B$, and $g_i$ the singular values of $\GG_{0,AB}$, then the resulting bound is given by~\cite{Molesky2020c}:
\begin{align}
   \Phi(\omega) \leq \sum_i \left[\frac{1}{2\pi}\Theta(\zeta_A \zeta_B g_i^2 - 1) + \frac{2}{\pi} \frac{\zeta_A \zeta_B g_i^2}{(1+\zeta_A\zeta_B g_i^2)^2} \Theta(1 - \zeta_A \zeta_B g_i^2) \right],
   \label{eq:NFRHTSFCB}
\end{align}
where $\zeta_{A,B} = |\chi_{A,B}|^2 / \Im \chi_{A,B}$. One can see that the expression of \eqref{NFRHTSFCB} has components of both material response (in $\zeta_{A,B}$) and channels (in the $g_i$ factors) in it. Strikingly, in the near-field\index{near field} limit, expression \eqref{NFRHTSFCB} is given by~\cite{Venkataram2020}
\begin{align}
    \Phi(\omega) \frac{d^2}{A} \leq &\frac{1}{4\pi^2} \ln\left(1 + \frac{\zeta_A \zeta_B}{4}\right) \nonumber \\
    &+ \frac{\Theta(\zeta_A \zeta_B - 4)}{8\pi^2}  \left\{ \ln(\zeta_A\zeta_B) + \frac{1}{4}\left[ \ln \left( \frac{\zeta_A \zeta_B}{4} \right) \right]^2 - 2 \ln\left(1 + \frac{\zeta_A \zeta_B}{4}\right) \right\},
   \label{eq:NFRHTSFCB2}
\end{align}
which correctly captures the \emph{logarithmic} material dependence that is seen in planar bodies. This significantly tightens the bound of \eqref{NFRHTSF} for plasmonic\index{plasmons} materials such as silver or gold which have large material enhancement factors\index{material enhancement factor} $|\chi|^2 / \Im \chi$. The genesis and utility of the bounds of \eqrefrange{NFRHTSF}{NFRHTSFCB2} are illustrated in \figref{SpectralNFRHT}, which contains the derivation of the conservation-law bounds of \eqref{NFRHTSF} in \figref{SpectralNFRHT}(a), the design of structures showing the material dependence of \eqref{NFRHTSF} in \figref{SpectralNFRHT}(b), and the more general combination of conservation law and channel-decomposition approach of \eqref{NFRHTSFCB2} in \figref{SpectralNFRHT}(c).

Generically, it is not possible to find ``tighter'' single-frequency dependencies than those that arise in \eqref{NFRHTSFCB2}, as both the distance and material-enhancement dependencies are achievable in realistic-material planar designs. The only possible improvements are the coefficient prefactors, as well as the correct material dependence away from the surface-plasmon\index{plasmons} frequency, suggesting that \eqref{NFRHTSFCB2} indeed captures the key tradeoffs in single-frequency NFRHT\index{near-field radiative heat transfer (NFRHT)}. A key remaining question, then, is what is possible over a broad bandwidth? This question is resolved in \secref{finiteband}.

\subsection{All-frequency sum rules}
\label{sec:allfreqsumrules}
In \secref{sumrules}, we developed the key elements need for sum rules\index{sum rules}: a causal linear response function, an objective that does not involve the conjugate of that function, and certain technical conditions (e.g. sufficient decay). Optical extinction is the prototype example, as the optical theorem\index{optical theorem} prescribes that extinction be proportional to the imaginary part of the overlap of the incident field with the induced polarization field, a quantity that is analytic (for plane-wave incident fields) in the upper-half plane. Within the past few years~\cite{sanders_manjavacas_2018,Shim2019}, it has been realized that there is a near-field\index{near field} analog of extinction: the local density of states, or LDOS\index{local density of states (LDOS)}. As derived in \secref{LDOS}, (electric) LDOS is given by the trace of the imaginary part of the (electric) Green's function, evaluated at the source location:
\begin{align}
    \textrm{LDOS}(\xv,\omega) = \Im \Tr \left[\frac{1}{\pi\omega} \GG(\xv,\xv,\omega) \right].
    \label{eq:LDOST}
\end{align}
The key similarity with extinction is that LDOS\index{local density of states (LDOS)} is the imaginary part of an amplitude, rather than a squared norm (which depends on the complex conjugate of that amplitude). At first blush, then, it would appear that one can port exactly the derivation used for extinction to derive sum rules\index{sum rules} for LDOS. However, there are three obstacles that must be overcome.

First, LDOS diverges at high frequencies. Ignoring the effects of a scatterer (which are effectively infinitely far away at infinitely large frequencies), and as seen below \eqref{LDOSe}, the free-space photon density of states scales as $\omega^2$ as frequency goes to infinity. A diverging LDOS violates the asymptotic-decay requirement of KK relations\index{Kramers--Kronig (KK) relations}, prohibiting a sum rule\index{sum rules}. The resolution, however, is straightforward: one should subtract the free-space LDOS $\rho_0(\omega)$ from the total LDOS, leaving only the scatterer-based contribution $\rho_s(\omega)$:
\begin{align}
    \rho_s(\xv,\omega) = \rho(\xv,\omega) - \rho_0(\omega) &= \Im \Tr \left[\frac{1}{\pi\omega} \left(\GG(\xv,\xv,\omega) - \GG_0(\xv,\xv,\omega) \right) \right] \nonumber \\
    &= \Im \Tr \left[\frac{1}{\pi\omega} \GG_s(\xv,\xv,\omega) \right],
\end{align}
where we define $\GG_s$ as the scattered-field part of the Green's function. After isolating the scatterer's contribution to the LDOS, one can verify that the ``scattered LDOS'' indeed decays sufficiently quickly at high frequencies~\cite{Shim2019}. Hence this approach of subtracting the free-space LDOS, an approach generalized in ``dispersion relations with one subtraction''~\cite{Nussenzveig1972}, resolves the first issue of diverging LDOS.

The second issue is that one is not free to arbitrarily choose the pole frequency for a KK relation\index{Kramers--Kronig (KK) relations} involving the scattered LDOS. The Green's function itself is finite and generically nonzero at every real frequency, but by definition the LDOS includes a factor of $1/\omega$, as in \eqref{LDOST}. (This does not correspond to a divergent LDOS at zero frequency, as the \emph{imaginary} part of the Green's function goes to zero at frequency, but the real part does not generically go to 0.) This function, then, already has a pole at the origin. One could try to move the pole to infinite frequency, for example by multiplying by $\omega / (\omega - \omega_0)$ and taking the limit as $\omega_0 \rightarrow \infty$, but the high-frequency asymptotic behavior of LDOS is quite complicated. Hence, there is likely only a single meaningful sum rule\index{sum rules} for near-field\index{near field} LDOS, which arises from the intrinsic pole at zero frequency.

The third issue is that the real part of the Green's function diverges, since the source and measurement locations coincide; sum rules\index{sum rules} relate the integral of the imaginary part to the real part (or vice versa), which leads to the impermissible evaluation of an infinite quantity. (Such an integral \emph{should} diverge; the free-space LDOS increases with frequency, meaning that any integral over all frequencies will of course diverge.) One resolution to this issue was proposed in \citeasnoun{barnett_loudon_1996}: to remove the longitudinal contribution to the Green's function, which removes the singularity and suggests that over all frequencies there can be no net change in spontaneous-emission enhancements\index{spontaneous emission}. But this removal thereby precludes the possibility for near-to-far-field coupling that is crucial for spontaneous-emission engineering\index{spontaneous emission}, which is why a conventional refractive-index sum rule\index{sum rules} is recovered. Instead, it was recognized in Refs.~\cite{sanders_manjavacas_2018,Shim2019} that there is an alternative mechanism for overcoming this obstacle: to subtract out the free-space LDOS term from the total term. The free-space term is the one responsible for the diverging real part, yet the free-space LDOS is exactly known and hence there is no need for a KK relation\index{Kramers--Kronig (KK) relations} for that part anyhow. Hence this obstacle is resolved by the same procedure as the first one, and we can proceed to deriving a scattered-LDOS\index{local density of states (LDOS)} sum rule\index{sum rules}.

\begin{figure*}[tb]
    \includegraphics[width=1\textwidth]{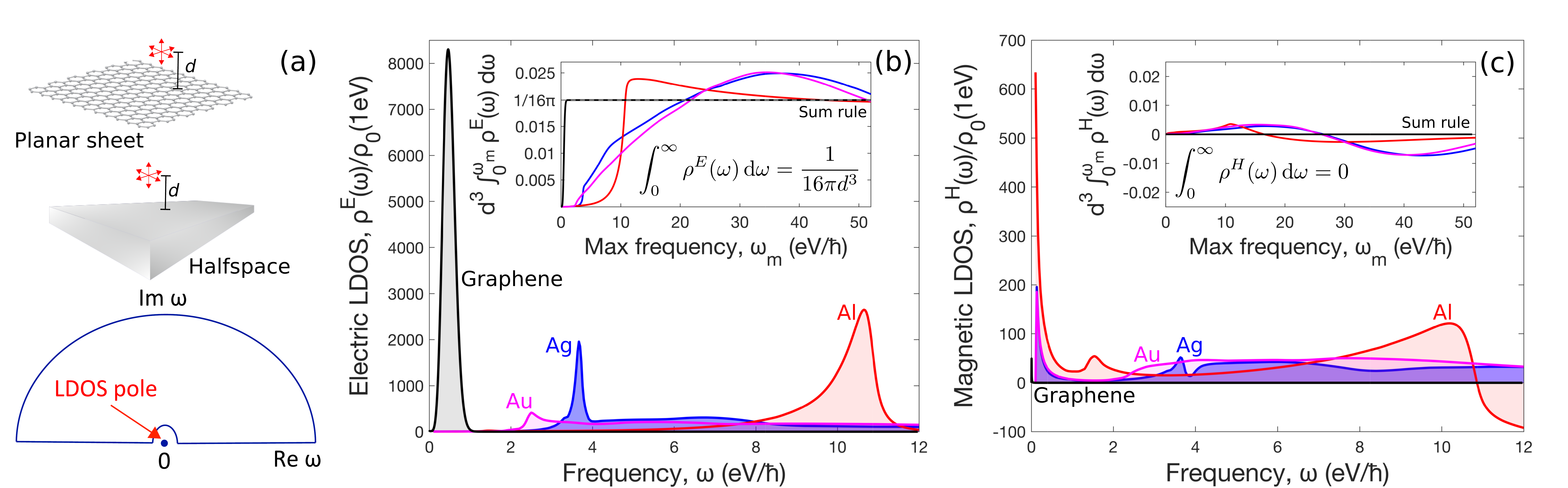}
    \caption{(a) Sum rules, derived using the techniques of \secref{sumrules} and the contour on the lower left, impose strong constraints on LDOS\index{local density of states (LDOS)} lineshapes. (b) Electric LDOS of various material half-spaces and 2D sheets, with different resonance peaks and bandwidths. The inset, however, shows that the integral converges to identical values for each scenario. (c) Similarly with magnetic LDOS, whose sum rule\index{sum rules} is now zero. The sum rules\index{sum rules} are for the scattered-field contributions to the LDOS, which can be negative at frequencies where spontaneous emission\index{spontaneous emission} is suppressed by the presence of a scatterer. (Adapted from \citeasnoun{Shim2019}.)}
    \label{fig:ldossumrule} 
\end{figure*} 

The hemispherical contour (with hemispherical bump at the origin), in tandem with the same Cauchy-residue\index{Cauchy residue theorem} arguments for far-field sum rules\index{sum rules} in \secref{sumrules}, leads to a sum rule\index{sum rules} for $\rho - \rho_0$ analogous to the far-field case~\cite{Shim2019}:
\begin{align}
    \int_0^\infty \rho_s(\omega,\xv) \dw = \frac{1}{2} \Re \Tr \GG_s(\xv,\xv) \big\vert_{\omega=0} = \alpha_{\rm LDOS}.
\end{align}
Now we have connected the all-frequency scattered-field component of electric LDOS to its electrostatic Green's function. Is that informative? It turns out to be quite informative, because there are near-field\index{near field} ``domain monotonicity'' theorems~\cite{Shim2019} that ensure that this shape-dependent Green's-function term is bounded above by its form in any enclosure, and we can choose high-symmetry enclosures where it has a simple analytical form. For example, for a planar half-space, the near-field\index{near field} electrostatic constant is simply 
\begin{align}
    \alpha_{\rm LDOS,plane} = \frac{1}{16\pi d^3} \left[ \frac{\varepsilon(0)-1}{\varepsilon(0)+1}\right],
    \label{eq:alphaplane}
\end{align}
where $\varepsilon(0)$ is the zero-frequency (electrostatic) permittivity. For conductive materials whose permittivity diverges at zero frequency, the corresponding fraction in \eqref{alphaplane} is simply 1, which can also be used as a general bound for any material. Notably, for the \emph{magnetic} LDOS\index{local density of states (LDOS)} above an electric material, the right-hand side of the counterpart to \eqref{alphaplane} is \emph{zero}: the scattering contribution to the magnetic LDOS must average out to zero (i.e., it provides suppression and enhancement of the free-space LDOS in equal amounts). 

An example of the utility of the LDOS sum rule\index{sum rules} is given in \figref{ldossumrule}. The electric LDOS is shown for three typical metals: gold (Au), silver (Ag), and aluminum (Al), as well as for a single graphene sheet (with Fermi level \SI{0.6}{eV}). These four systems show LDOS peaks at quite different frequencies, from below \SI{1}{eV} to beyond \SI{10}{eV}, with very different quality factors leading to quite different ``spreads'' in their spectral response. Yet as is made clear by the inset of \figref{ldossumrule}, the integrated response is exactly equal for each of these systems, as must be true from \eqref{alphaplane} (the material constant $\alpha$ for each system is exactly 1). Sum rules illuminate unifying principles that must apply across seemingly disparate systems.

\subsection{Finite, nonzero bandwidth}
\label{sec:finiteband}
The techniques of the previous two sections apply to single-frequency and all-frequency scenarios. In this section, we probe an intermediate regime: finite, nonzero bandwidth. Techniques that work for any arbitrary bandwidth would be tantalizingly powerful, as they would incorporate the single- and all-frequency results as asymptotic limits of a more general theory. Yet the techinques of the previous section would seem incapable of extension to nonnzero, finite bandwidths: there is no single scattering problem for which power-conservation laws can be imposed, nor can the contour integrals of the sum-rule\index{sum rules} approaches be easily modified to a finite bandwidth. In this section, we describe two recently developed approaches to tackle finite-bandwidth bounds: first, transforming bandwidth-averaged response to a complex frequency (largely following \citeasnoun{Shim2019}), and second, identifying an oscillator-based representation of any scattering matrix (largely following \citeasnoun{Zhang2022}).

\subsubsection{Complex-frequency bounds}
\label{sec:complexfreqbounds}
\citeasnoun{Shim2019} recognized an intermediate route that utilized both techniques in one fell swoop. The idea can be summarized succinctly: finite-bandwidth average response can be transformed to a scattering problem at a single, \emph{complex}-valued frequency, where quadratic constraints analogous to power conservation can be imposed. The complex frequency accounts for bandwidth, while the power-conservation analog imposes a finite bound. We now develop this intuition mathematically.

To compute the bandwidth average of a response function such as LDOS\index{local density of states (LDOS)}, one must define a ``window function'' that encodes the center frequency, the bandwidth, and the nature of the averaging. A common choice is a linear combination of step functions, but this choice turns out to be mathematically treacherous. A simple (and mathematically serendipitous) choice is a Lorentzian function. Uses of tailored window functions for bandwidth averaging were first proposed in Refs.~\cite{hashemi_qiu_mccauley_joannopoulos_johnson_2012,Liang2013}; in the first, bandwidth-averaged extinction was analyzed for scaling laws for optical cloaking, while in the second, they were used to regularize the computational inverse design of maximum LDOS. Our quantity of interest, the frequency-averaged LDOS, $\langle \rho \rangle$, can be written~\cite{Shim2019}
\begin{align}
    \langle \rho \rangle = \int_{-\infty}^{\infty} \rho(\omega) H_{\omega_0,\Delta\omega}(\omega) \,{\rm d}\omega,
    \label{eq:rhoavg}
\end{align}
where $H_{\omega_0,\Delta\omega}(\omega)$ is the Lorentzian window function,
\begin{align}
    H_{\omega_0,\Delta\omega}(\omega) = \frac{\Delta \omega / \pi}{(\omega - \omega_0)^2 + (\Delta\omega)^2},
\end{align}
where $\omega_0$ is the center frequency and $\Delta \omega$ is the bandwidth of interest. In \eqref{rhoavg} we define the frequency integral from $-\infty$ instead of 0 for smoothness; typically, the window function will be narrow enough to render this difference negligible; conversely, in the all-frequency limit, the symmetry of the LDOS around zero frequency ensures we are working with the correct quantity. We are interested only in the near-field\index{near field} enhancements of $\rho$, so we will drop the free-space LDOS, as was useful in the sum-rule\index{sum rules} section to avoid spatial and spectral divergences. Then, consider the integral of \eqref{rhoavg}: it already covers the entire real line, we can imagine adding to it a the hemispherical contour in the UHP that will contribute infinitesimally. Then the integral is a closed contour, and we can use complex-analytic techniques based on the analyticity of the integrand and the locations of the poles of the integrand. The integrand is not analytic, but the LDOS can be written as $\rho(\omega) = \Im s(\omega)$, where $s(\omega)$, proportional to the trace of the imaginary part of the scattered component of the Green's function, \emph{is} analytic. Taking the imaginary part outside the integral, the remainder of the integrand of \eqref{rhoavg} has two poles away from the lower-half plane: one at zero, thanks to the $1/\omega$ term in the LDOS, and a second at $\omega_0 + i\Delta\omega$. Then, a few lines of algebra gives the frequency average of $\rho(\omega)$ as~\cite{Shim2019}
\begin{align}
    \langle \rho \rangle = \Im s(\omega_0 + i\Delta\omega) + 2 H_{\omega_0,\Delta\omega}(0)\alpha_{\rm LDOS}.
\end{align}
The second term comes from the contribution of the sum rule\index{sum rules} at a given frequency, and ensures that the ultimate expression will give the sum rule\index{sum rules} in the asymptotic limit $\Delta\omega \rightarrow \infty$. Here, for simplicity and pedagogy, we will assume a sufficiently narrow bandwidth that the second term can be ignored. (It can always be reintroduced in the final expression.) The first term is the imaginary part of the LDOS\index{local density of states (LDOS)} scattering amplitude, evaluated at the \emph{complex} frequency $\wt = \omega_0 + i\Delta\omega$. What is the largest this term can be?

To bound the complex-frequency term, we can develop a generalization of the real-frequency conservation-law approach. In \citeasnoun{Shim2019} we developed such a generalization via a somewhat complicated line of differential-equation reasoning; here, we develop a simpler (but no less general) integral-equation form. The starting point is the complex-valued integral equation,
\begin{align}
    \left[\GG_0(\wt) + \xi(\wt)\right] \pv(\wt) = -\einc(\wt),
    \label{eq:VIEcomplex}
\end{align}
where we have momentarily included all frequency arguments to emphasize that \eqref{VIEcomplex} is evaluated at the complex frequency $\wt$. Next, we will multiply on the left by $\pv^\dagger / \wt$, and take the imaginary part of the entire equation, to arrive at 
\begin{align}
    \pv^\dagger \left\{\Im \left[\frac{\GG_0}{\wt} + \frac{\xi}{\wt}\right] \right\} \pv = \Im \left[\left(\frac{\einc}{\wt}\right)^\dagger \pv\right],
    \label{eq:optthmcomplex}
\end{align}
This equation can be regarded as a complex-valued extension of the real-valued, global conservation law\index{global conservation laws} of \eqref{optthm}. In particular, the two terms on the left are both positive-semidefinite, as can be proven by causality\index{causality} (cf. Sec.~IX of the SM of \citeasnoun{Kuang2020}). To remove the shape dependence and focus on the material dependence, then, we can drop the first term on the left-hand side of \eqref{optthmcomplex}, and rewrite this equation as an inequality:
\begin{align}
    \pv^\dagger \left[\Im \left(\frac{\xi}{\wt}\right) \right] \pv \leq \Im \left[\left(\frac{\einc}{\wt}\right)^\dagger \pv\right],
    \label{eq:optthmcomplex2}
\end{align}
\Eqref{optthmcomplex2} imposes a constraint on the strength of the complex-frequency polarization field that enters the near-field\index{near field} scattering amplitude $s(\wt)$. The exact expression for the scattering amplitude is $s(\wt) = \frac{1}{\pi\wt} \Tr \GG_0(\xv,\xv,\wt)$. One can maximize the imaginary part of this amplitude subject to the constraint of \eqref{optthmcomplex2} by exactly the procedure outlined in Sec.~IX of the SM of \citeasnoun{Shim2019}; doing so, one arrives at a simple result (remembering that we have dropped the sum-rule\index{sum rules} term):
\begin{align}
    \langle \rho \rangle \leq \frac{1}{\pi} \frac{|\chi(\wt)|^2}{\Im[\wt\chi(\wt)]} \einc^\dagger \einc.
\end{align}
As a reminder, the inner product of the incident field with itself is a volume integral of the square of the incident fields. The deep near field\index{near field} is dominated by the most rapidly decaying term in the incident fields; integrating only this contribution at the complex frequency gives $\einc^\dagger \einc = \frac{1}{16\pi d^3}$, where we have taken the arbitrary scattering body to fit in a halfspace enclosure separated from the source by a distance $d$. Inserting this expression into the inequality, and normalizing by the free-space LDOS evaluated at $|\wt|$, we finally have a bandwidth-averaged bound~\cite{Shim2019}:
\begin{align}
    \frac{\langle \rho \rangle}{\rho_0(|\wt|)} \leq \frac{1}{8|k|^3 d^3} f(\omega),
    \label{eq:rhoavgbnd}
\end{align}
where $f(\omega)$ is the bandwidth-averaged generalization of the material-enhancement factor (discussed at real frequencies in \secref{SEbounds}),
\begin{align}
    f(\omega) = \frac{|\wt \chi|^2}{|\wt| \Im\left(\wt \chi\right)}.
    \label{eq:fcomplex}
\end{align}
The material enhancement factor\index{material enhancement factor} of \eqref{fcomplex} is slightly simpler than that of \citeasnoun{Shim2019}, thanks to our use of the simpler integral-equation constraint of \eqref{optthmcomplex}.

The bound of \eqref{rhoavgbnd} is the key result: the bandwidth-averaged LDOS\index{local density of states (LDOS)} has an upper bound that is similar to that of the single-frequency LDOS, but reduced by the presence of a complex frequency. This reduction is significant for low-loss materials, for which $\Im \chi$ might be quite small, in which case $\Im(\wt\chi) \approx (\Delta \omega) \chi$, wherein the bandwidth effectively provides the relevant loss. There is also an additional broadening due to dispersion, as $\chi$ is evaluated at the complex frequency $\wt$, at which $\Im \chi$ will generally be larger. (There is \emph{another} additional term in the more general version of the bound of \eqref{rhoavgbnd} that exponentially decays with bandwidth, that we excluded for simplicity.) Hence the bound of \eqref{rhoavgbnd} has three properties that are quite theoretically pleasing. First, in the single-frequency limit, it asymptotically approaches the previously derived single-frequency bound. Second, in the all-frequency limit, it asymptotically approaches the previously derived sum rule\index{sum rules}. And, finally, in the nonzero- and finite-bandwidth regime, it intermediates between the two, with a smaller average response than the single-frequency bound, and a smaller total integrated response than the sum rule\index{sum rules}. This approach was extended to CDOS\index{cross density of states (CDOS)} and NFRHT\index{near-field radiative heat transfer (NFRHT)} as well in \citeasnoun{Shim2019}, with similar features emerging. One interesting comparison point is to \citeasnoun{Zhang2020}, which examined optimal materials for planar NFRHT\index{near-field radiative heat transfer (NFRHT)} designs. Unlike the power--bandwidth bounds, which increase with electron density and decrease with material loss, \citeasnoun{Zhang2020} found that the key material parameters in planar systems are simply the (ideally small) frequency at which surface polaritons\index{polaritons} are strongest, and the bandwidth over which they are strong. This finding has been experimentally corroborated~\cite{Mittapally2023}, and it emerges theoretically in the more general NFRHT\index{near-field radiative heat transfer (NFRHT)} bounds of the next subsection.

\citeasnoun{Shim2019} probed the feasibility of approaching the upper bounds in certain prototypical systems. Four key results were identified. First, for center frequencies close to the surface-plasmon\index{plasmons} frequencies of metals, planar systems supporting such plasmons\index{plasmons} are able to closely approach the bounds across a wide range of bandwidths. Second, double-cone (bowtie-antenna-like) antennas show a performance that can closely approach (nearly within 2X) their bounds across a wide range of bandwidths, for center frequencies coincident with their resonant frequencies. Third, these bounds were the first to enable systematic comparison of dielectric- and metal-based systems. Unlike the single-frequency case, the complex-frequency material enhancement factor\index{material enhancement factor} does not diverge for lossless dielectrics (at nonzero bandwidth), which enables predictions of the center frequencies and bandwidths at which metals can be categorically superior to dielectrics, and vice versa. Finally, these bounds also enabled predictions of when 2D materials can be superior to bulk materials, and vice versa. The results highlight the power of fundamental limits more generally: they enable a high-level understanding of the landscape of a given physical design problem, identifying the material and architectural properties that really matter.

The ``power--bandwidth'' approach of \citeasnoun{Shim2019} was recently generalized in \citeasnoun{Chao2022} to incorporate the concept of local conservation laws\index{local conservation laws} into the picture. Notice that the constraint of \eqref{optthmcomplex} is a global conservation law\index{global conservation laws}; at the time that \citeasnoun{Shim2019} was published, the local-conservation-law\index{local conservation laws} approach had not yet been invented. \citeasnoun{Chao2022} remedies this gap, and shows that for dielectric scatterers, the use of additional conservation laws can significantly improve the resulting bounds. There is an interesting interplay between the quality factor of the sources and the bandwidth of interest, and there are useful semi-analytical bounds that can be derived from the global conservation laws applied to large-scale devices. Moreover, inverse-design structures are shown to come quite close to the improved complex-frequency, local-conservation-law\index{local conservation laws} bounds.

\subsubsection{Oscillator-representation bounds}
An alternative to the complex-frequency approach to bandwidth averaging was very recently proposed in \citeasnoun{Zhang2022}. We will briefly summarize the (detailed) mathematical apparatus developed, and highlight the key result for our purposes: a new, nearly tight bound for bandwidth-averaged NFRHT\index{near-field radiative heat transfer (NFRHT)}.

Before delving into scattering bodies, consider the bulk optical susceptibility of a material. It is known that the response of an isotropic passive material\index{passivity} can be written as a linear combination of Drude--Lorentz oscillators,
\begin{align}
    \chi(\omega) = \sum_i \frac{\omega_p^2}{\omega_i^2 - \omega^2 - i\gamma\omega} c_i,
    \label{eq:chiDL}
\end{align}
where $\omega_p$ is the ``plasma frequency'' of the material (related to its electron density~\cite{Yang2015,Kaxiras2019}), $\omega_i$ are the oscillator frequencies, $\gamma$ are infinitesimal oscillator loss rates, and the $c_i$ are ``oscillator strengths'' that sum to unity thanks to the sum rule\index{sum rules} of \eqref{chiSumInf} discussed in \secref{sumrules}. Often this representation is derived in single-electron quantum-material frameworks~\cite{Kaxiras2019}, but it applies more generally as a consequence of causality\index{causality} and passivity\index{passivity}. (The technically rigorous mathematical statement uses the theory of Herglotz functions~\cite{Bernland2011}.) Any linear material's susceptibility must conform to the Drude-Lorentz linear combination of \eqref{chiDL}; perhaps not with a small number of oscillators (it is well known that effects such as inhomogeneous broadening lead to other lineshapes, such as the ``Voigt'' lineshape~\cite{Hartmann2021}), but with sufficiently many oscillators. It may seem counter-intuitive to work with a representation that may need 1,000, or even 100,000 oscillators, instead of a different model with fewer parameters. From an optimization perspective, however, this is not correct. In the Drude--Lorentz representation of \eqref{chiDL}, the only degrees of freedom are the $c_i$ coefficients, and the susceptibility is \emph{linear} in these degrees of freedom. In many scenarios, large linear optimization problems are significantly easier to solve (sometimes even analytically) than large, nonlinear (and nonconvex) optimization problems.

Cauality\index{causality} and passivity\index{passivity} create three key ingredients that together lead to the Drude--Lorentz representation of \eqref{chiDL}: a Kramers--Kronig relation\index{Kramers--Kronig (KK) relations}, a sum rule\index{sum rules}, and positivity of the imaginary part of the susceptibility. The exact sequence of transforming those ingredients to the Drude--Lorentz representation is detailed in \citeasnoun{Shim2021}. One intuitive description is that the imaginary part of the susceptibility is a positive quantity, and can be discretized into coefficients at many discrete frequencies along the real axis. Passivity\index{passivity} implies that these coefficients are real, while the sum rule\index{sum rules} implies that their sum is constrained. Finally, the Kramers--Kronig relation\index{Kramers--Kronig (KK) relations} guarantees that the imaginary parts of the susceptibilities are the \emph{only} degrees of freedom; the real parts are entirely determined by the imaginary parts. Compiling the mathematical details of these steps leads to \eqref{chiDL}, which is a relation that many find intuitive thanks largely to the fact that it can be derived in single-electron quantum mechanics.

The key idea of \citeasnoun{Zhang2022} is that there is a \emph{wave-scattering operator} that exhibits nearly identical mathematical properties to material susceptibilities. This operator is the ``$\TT$'' matrix. The $\TT$ matrix is a scattering matrix that relates the polarization field induced in any scattering body to the incident fields impinging upong it~\cite{Carminati2021}:
\begin{align}
    \Pv(\xv,\omega) = \int_V \TT(\xv,\xv',\omega) \Einc(\xv',\omega) \,{\rm d}\xv',
\end{align}
or, in vector notation:
\begin{align}
    \pv = \TT \einc.
\end{align}
The $\TT$ matrix is a \emph{causal\index{causality} linear response function}, as the polarization field at $\xv$ cannot be excited before the incident field exciting it reaches $\xv'$. Just as causality\index{causality} implies a Kramers--Kronig relation\index{Kramers--Kronig (KK) relations} for material susceptibilities, it was recognized in \citeasnoun{Zhang2022} that causality\index{causality} implies a Kramers--Kronig relation\index{Kramers--Kronig (KK) relations} for $\TT$ matrices. Sum rules come from the low- and high-frequency asymptotic behavior of Kramers--Kronig relations\index{Kramers--Kronig (KK) relations}, and the $\TT$ matrix satisfies a matrix-valued analog of the $f$-sum rule\index{sum rules} for material oscillator strengths. Finally, just as passivity\index{passivity} implies that the imaginary parts of susceptibilities are positive, it similarly implies that the anti-Hermitian part of the $\TT$ matrix is positive semidefinite. Together, these three ingredients imply a matrix-valued analog of \eqref{chiDL} for any $\TT$ matrix:
\begin{align}
    \TT(\omega) = \sum_i \frac{\omega_p^2}{\omega_i^2 - \omega^2 - i\gamma\omega} \TT_i,
    \label{eq:TDL}
\end{align}
where the Drude--Lorentz parameters are exactly the same as in \eqref{chiDL}, and the $\TT_i$ are now matrix-valued coefficient degrees of freedom. The exact expression of \eqref{TDL} is for the case of reciprocal materials; in nonreciprocal terms there is an extra term that makes the calculations more tedious but has no effect on most applications of interest. Analogous to the constraints on material oscillator strengths, passivity\index{passivity} and the $\TT$-matrix sum rule\index{sum rules} lead to constraints on the $\TT_i$:
\begin{align}
    \sum_i \TT_i = \II, \qquad \TT_i \geq 0,
    \label{eq:TTconstraints}
\end{align}
where $\II$ is the identity matrix. \Eqref{TDL}, and its nonreciprocal analog, must hold for \emph{any} linear electromagnetic scattering process. Even in scattering processes with complex interference phenomena, Fano resonances, etc., $\TT(\omega)$ must exhibit lineshapes consistent with \eqref{TDL}, which is shown in \citeasnoun{Zhang2022} to reveal surprising structure even in typical scattering problems.

Our interest in this chapter, however, is in fundamental limits, so we will focus on the utility of \eqref{TDL} to identify upper bounds in the application considered in \citeasnoun{Zhang2022}, which is NFRHT\index{near-field radiative heat transfer (NFRHT)}. The approach in the paper requires a dozen or so mathematical steps explained in Sec.~IX of the SM of \citeasnoun{Zhang2022}; the key is to transform the problem from one of thermal sources inside the hot body radiating power to the cold one to one of incoherent sources \emph{between} the bodies radiating back to the emitter body. There are various other key steps, such as an appropriate renormalization of the point sources between the bodies. Ultimately, the culmination is the following: NFRHT\index{near-field radiative heat transfer (NFRHT)} is rewritten in terms of the total $\TT$ matrix of the collective bodies, at which point the representation of \eqref{TDL} is inserted. Then, the entire frequency dependence of the problem is given by the collective products of the Drude--Lorentz oscillators and the Planck function, whose integrals can be determined analytically. Then one is left with a linear summation of given coefficients multiplying the unknown $\TT_i$ degrees of freedom. The optimization over all possible $\TT_i$, subject to the constraints of \eqref{TTconstraints}, has many unknowns, but can be done \emph{analytically}, leading to a simple yet completely general bound on thermal HTC\index{heat transfer coefficient (HTC)}:
\begin{align}
    \textrm{HTC} \leq \beta \frac{T}{d^2},
    \label{eq:HTCbound}
\end{align}
where $T$ is the temperature, $d$ is the separation, and $\beta \approx 0.11 k_B^2 / \hbar$ is a numerical constant. \Eqref{HTCbound} is an unsurpassable limit that captures the key constraints imposed on every scattering $\TT$ matrix. Strikingly, despite the relative simplicity of the approach, it offers the tightest bounds on NFRHT\index{near-field radiative heat transfer (NFRHT)} to date, only a factor of 5 larger than the best theoretical designs~\cite{Zhang2020}. Previous approaches suggested strong material dependencies, with bounds that increased with electron density, whereas planar designs show the reverse trend. In this bound, use of a low-frequency sum rule\index{sum rules} in the $\TT$-matrix representation leads to an electron-density-independent bound. Moreover, the optimization over $\TT_i$ predicts precisely the same optimal peak transfer frequency as the best designs~\cite{Zhang2022}. 

There are two sets of relaxations used to arrive at the bound of \eqref{HTCbound}: first, beyond the representation theorem, no other Maxwell-equation constraints are imposed. Hence the optimal $\TT_i$ may not actually be physically realizable. Potentially one could impose such constraints exactly by the local-conservation-law approach\index{local conservation laws} discussed above. Second, the heat transfer process is relaxed to the emission of the sources between the bodies into both the source and emitter, whereas the exact expression is the \emph{difference} between the radiation into the emitter and receiver bodies. The latter relaxation leads to a linear dependence on $\TT(\omega)$, as opposed to the qudaratic dependence in the exact expression. It may be possible to optimize over the exact quadratic expression using manifold optimization techniques~\cite{Absil2009,manopt,Zhang2022}. Tightening these relaxations may lead to a further tightening of the bound. Conversely, they may lead to the same bound, and improved design techniques~\cite{Gertler2023subm} may identify structures that can achieve them.

\subsection{Mode volume}
\label{sec:modevbounds}
In this final section, we turn to the question of bounds on mode volume\index{mode volume}. Mode volume is a very different response function than any of those previously considered, as it is a property of an eigenfunction rather than a scattering quantity. There is no incident field in the definition of a mode volume\index{mode volume}, and hence the power-conservation and causality-based\index{causality} approaches of the previous sections are not immediately useful. In this section, we describe a method for bounding minimum mode volumes\index{mode volume} based on the optimization-theoretic notion of duality\index{dual problem}.

In optimization theory, the \textbf{dual}\index{dual problem} of an optimization problem is a second optimization problem, related to but distinct from the original, ``primal'' optimization problem~\cite{Boyd2004}. The dual problem\index{dual problem} is formed by incorporating all constraints into the Lagrangian of the original optimization problem, introducing Lagrange multipliers as coefficients of the constraints, and optimizing out the primal variables, leaving only the Lagrange multipliers as degrees of freedom. An equivalent interpretation is that if one interprets a generic minimization optimization problem as the \emph{minimax} of a Lagrangian, the dual problem\index{dual problem} is the \emph{maximin} of the same Lagrangian. The dual program\index{dual problem} has two properties that can be quite useful for optimization and bounds: it is always a concave maximization problem (equivalent to a convex minimization problem, and therefore efficiently solvable by standard convex-optimization techniques), and its maximum is guaranteed to be a lower bound for the orginal, primal, minimization problem. 

For many optimization problems, the dual\index{dual problem} cannot be expressed in a simple form; even amongst those problems for which it has a simple expression, it often has the trivial solution $-\infty$ as its maximum, giving a trivial lower bound. \citeasnoun{Angeris2019} showed that a very special class of electromagnetic design problems have a nontrivial, semi-analytical dual problem\index{dual problem}. In particular, for design problems in which the objective function to be minimized is the norm of a difference between the electric field $\Ev$ and some target field $\Ev_{\rm target}$,
\begin{align}
    \mathcal{F} = \| \Ev - \Ev_{\rm target} \|^2,
    \label{eq:Ftar}
\end{align}
then one can impose the full Maxwell-equation constraints and identify a non-trivial, semi-analytical dual problem\index{dual problem}. One might suspect that objectives of the form of \eqref{Ftar} might be quite common: after all, a focusing metalens could have a target field that matches an Airy beam along a focal plane, a surface-pattern design intended to maximixe spontaneous-emission\index{spontaneous emission} enhancements could target the field at the location of the dipole, and so forth. But these cases do not work for the expression of \eqref{Ftar}: for a non-trivial dual problem\index{dual problem}, the field $\Ev_{\rm target}$ must be specified \emph{at every spatial point of the entire domain}. This includes, for examples, the points within the scatterer, the points within any PML regions, etc. Knowing a target field at a single point, or on a focal plane, is not sufficient. And it is hard to think of any application in which we know the target field across the entire domain.

It turns out, however, that mode-volume minimization can be reformulated to target an objective specified over the entire domain. Mode volume, as specified in \eqref{Vm}, is given by the integral of the field energy over all space divided by the field energy at a single point. Typically the integral is treated as a normalization constant (taken to equal 1), and maximization of the field energy at a single point is the key objective. In \citeasnoun{Zhao2020}, it was recognized that this convention could be reversed: the field energy at the point of interest can be fixed as a normalization constant, equal to 1, while minimizing the integral of the field energy can be the objective. Such an objective is exactly of the form of \eqref{Ftar}, with a target field of 0 everywhere! Physically, this makes intuitive sense: a minimum mode volume\index{mode volume} tries to minimize the field energy at every point, except for the ``origin'' of interest; everywhere else, it wants to drive the field as close to a target of 0 as possible.

Given this transformation, and a few others described in \citeasnoun{Zhao2020}, one can use the formulation of \citeasnoun{Angeris2019} to specify a dual program\index{dual problem} for the mode-volume minimization problem. The solutions of this dual program\index{dual problem} can be formulated with the modeling language CVX~\cite{cvx} and solved with Gurobi~\cite{gurobi}, and those solutions represents fundamental lower bounds on the mode volume\index{mode volume}, given only a designable region and a refractive index of the material to be patterned.

First, the 2D TE case encapsulates scalar-wave physics: without vector fields, there also are not the field discontinuities across boundaries that can be responsible for large field amplitudes in ``slot-mode'' configurations~\cite{Lipson05,Choi2017,Hu2018}. There also is no near field\index{near field} for scalar waves, in the sense of large nonpropagating fields that culminate in a singularity at the location of a point source. In this case, the argument for a trivially small mode volume\index{mode volume} near a perfectly sharp tip \emph{fails}: the lack of a singularity means that one cannot drive the field at the location of the source arbitrarily high. If there is to be no sharp-tip enhancement (as we will see), then dimensional arguments would require mode volume\index{mode volume} to scale with the square of the wavelength (in 2D), restoring the notion of a ``diffraction-limited'' mode volume\index{mode volume}. The only question, then, is the value of the coefficient of the squared wavelength. The duality-computed\index{dual problem} bounds confirm indeed that below some separation distance $d$, the mode-volume bounds asypmtotically flattens out, to a small fraction of the square wavelength. This bound depends only on the available refractive index of the designable region, and has been closely approached by inverse-designed structures~\cite{Liang2013,Zhao2020}.

The 2D TM case is fundamentally different: sharp field discontinuities occur across material boundaries, and singularities in the near field\index{near field} of point sources imply the possibility for zero mode volume\index{mode volume} unless fabrication constraints, or similarly a nonzero source--scatterer separation distance, is enforced. In this case, the duality-based\index{dual problem} approach finds quite different scaling: the 2D TM mode-volume bounds scale as $d^2$, where $d$ is the relevant source--scatterer distance (or sharp-tip radius of curvature), with no dependence on the wavelength. Intriguingly, this scaling is faster than the typical structure used for mode-volume minimization: a ``bowtie antenna''~\cite{Choi2017,Hu2018}, whose optimal mode volume\index{mode volume} appears to scale only linearly with $d$ (and hence linearly with wavelength, $\lambda$, as well). In \citeasnoun{Zhao2020}, it is shown that inverse-designed structures appear to exhibit mode volumes\index{mode volume} that scale roughly as $d^{1.4}$, faster than the linear scaling of bowtie antennas but not quite as fast as the duality-based\index{dual problem} bound. At smaller length scales, these differences can be dramatic. For minimum feature sizes $d \approx 0.01\lambda$, the inverse-design curve falls about 5X below the bowtie-antenna curve, which itself is 40X above the mode-volume bound. Resolving this gap, either through identifying better designs or by identifying tighter bounds, could lead to significant reductions in mode volume\index{mode volume} through near-field\index{near field} engineering.

\section{Summary and looking forward}
Near-field optical response can require significant mathematical machinery, and the techniques to bound them even moreso. We were careful above to give correct and sometimes nearly complete mathematical descriptions. Here, we can give a high-level summary of three of the prototypical response functions and application areas covered:
\begin{itemize}
    \item LDOS\index{local density of states (LDOS)}, arguably the most important near-field\index{near field} response function, has single-frequency bounds that scale as $1/d^2$ and $|\chi(\omega)|^2 / \Im \chi(\omega)$~\cite{Miller2016}. This bound can be achieved at the surface-plasmon\index{plasmons} frequency of a given material; away from that frequency, inverse designs have shown good performance that can be relatively close to the bound, but generally it is also true that tighter bounds can be computed by using additional constraints. A sum rule\index{sum rules} is known for all-frequency LDOS~\cite{sanders_manjavacas_2018,Shim2019}, which depends on the separation but \emph{not} on the material; over finite bandwidth, bounds similar to the single-frequency expression can be found, albeit evaluated at the complex frequency. Again, these bounds are nearly achievable when the frequency range is centered around the surface-plasmon\index{plasmons} frequency of a material, but can be tightened in other scenarios (e.g. dielectric materials)~  \cite{Molesky2020}. The key open questions around LDOS are two-fold: first, is there an analytical or semi-analytical bound that can be derived that is nearly achievable across all frequencies? And can one identify achievable bounds for only the \emph{radiative} part of the LDOS, i.e., that fraction of power that is emitted to the far field?
    \item Near-field radiative heat transfer is one of the most technically challenging areas of near-field\index{near field} optics, both experimentally and theoretically, but an abundance of work makes it perhaps the area where we have the best understanding of what is possible. For planar bodies, there are simple and powerful transmission expressions for NFRHT\index{near-field radiative heat transfer (NFRHT)}~\cite{Pendry1999,Biehs2010}, as well as an understanding of the optimal materials that lead to the largest response~\cite{Zhang2020,Rousseau2012,Pascale2023}. At a single frequency, semi-analytical bounds have been derived~\cite{Venkataram2020} that scale as $1/d^2$ with separation distance and \emph{logarithmically} with $|\chi(\omega)|^2 / \Im \chi$, both dependencies of which are exhibited by planar structures. Finally, when averaging against the Planck function to account for the thermal nature of the radiation, the recently developed oscillator theory of $\TT$ matrices~\cite{Zhang2022} enables a bound proportional only to $1/d^2$ and $k_B^2 T / \hbar$, with no material dependence. This bound can be approached within a factor of five by the best theoretical designs, showing a comprehensive understanding of what is possible in NFRHT\index{near-field radiative heat transfer (NFRHT)}, and the materials and structures needed to achieve that performance. One interesting open question is how this bound changes when one of the bodies must have a bandgap, as is required, for example, in thermophotovoltaics.
    \item Finally, mode volume\index{mode volume} is quite different from the other response functions considered above. It is a property of an eigenmode, instead of a scattered field, and hence some of the techniques based on power conservation do not lead to useful bounds in this case. The only approach we know of that leads to useful bounds relies on the \emph{duality}\index{dual problem} technique of optimization theory. The most important question surrounding mode volume\index{mode volume} is how it scales with minimum feature size $d$. Ideally, it would scale as $d^n$, where $n$ is the dimensionality of the system (either 2D or 3D), with no dependence on wavelength; this scaling would lead to the largest enhancements at highly subwavelength feature sizes. Certainly such scaling is possible with plasmonic\index{plasmons} structures, but plasmonic\index{plasmons} structures are too lossy, and the concept of mode volume\index{mode volume} itself must be modified for plasmonic\index{plasmons} mode volume\index{mode volume}~\cite{Lalanne2018}. The question, then, is the optimal scaling for dielectric materials. Interestingly, the duality-based\index{dual problem} bounds of \citeasnoun{Zhao2020} suggest exactly $d^n$ scaling. However, bowtie-antenna structures show $d^{n-1}$ scaling, while inverse designs appear to show a scaling between these two. Hence progress has been made on this crucial question, but it is still not fully resolved: what is the best possible scaling of mode volume\index{mode volume} with minimum feature size?
\end{itemize}

The theory of fundamental limits to near-field\index{near field} optical response is now sufficiently rich to be summarized in a book chapter, as we have done here. But the story is not complete: as we have seen in numerous examples, including the three above, there are still many response functions, material regimes, and frequency ranges at which there are gaps between the best known device structures and the best known bounds. Many of the bound techniques described herein have only been discovered in the past few years, and there are likely still significant strides to be made. The optical near field\index{near field} continues to offer a fertile playground for theoretical discovery, experimental demonstration, and new devices and technological applications.

\section{Appendix: Complex analysis for sum rules}
\label{sec:appCA}
Here we provide a brief summary of the basic rules of complex analysis, and how they are derived, emphasizing the key results relevant to sum rules\index{sum rules}. More expansive discussions of these ideas can be found in any good complex-analysis textbook.

First, we start with the definition of \textbf{complex differentiable}: a function $f$ is complex differentiable if the limit
\begin{align}
    f'(z) = \lim_{h\rightarrow 0} \frac{f(z+h) - f(z)}{h}
\end{align}
exists for $h$ along \emph{any path} in the complex plane. The equality along any path is a very strong constraint, and leads to the Cauchy--Riemann conditions on the derivatives of the real and imaginary parts of $f$. A function that is complex differentiable at every point on some domain $\Omega$ is \textbf{holomorphic} on $\Omega$. A major theorem of complex analysis is that all such functions are also \textbf{complex analytic} (which means they have a convergent power series in a neighborhood of every point in $\Omega$). From complex differentiability, it is a straight path to \textbf{Cauchy's integral theorem}: for $f$ holomorphic on $\Omega$, and a closed contour $\gamma$ in $\Omega$,
\begin{align}
    \oint_\gamma f(z) \,{\rm d}z = 0,
\end{align}
which can be proven by setting $f = u + iv$, ${\rm d}z = {\rm d}x + i{\rm d}y$, applying Green's / Stokes theorem, and using the Cauchy--Riemann conditions.

An important technique for integrals over open contours is \textbf{contour shifting}: if $\gamma$ and $\tilde{\gamma}$ are contours with the same endpoints, then
\begin{align}
    \int_\gamma f(z) \,{\rm d}z = \int_{\tilde{\gamma}} f(z) \,{\rm d}z.
\end{align}
This follows directly from reversing the second contour, combining it with the first to make a closed contour, and applying Cauchy's integral theorem. Contour shifting is common in Casimir physics, for example, where the standard transformation is a ``Wick rotation'' from the positive real axis to the positive imaginary axis~\cite{Johnson2011}.

\begin{figure}[htb]
    \centering\includegraphics[width=0.6\linewidth]{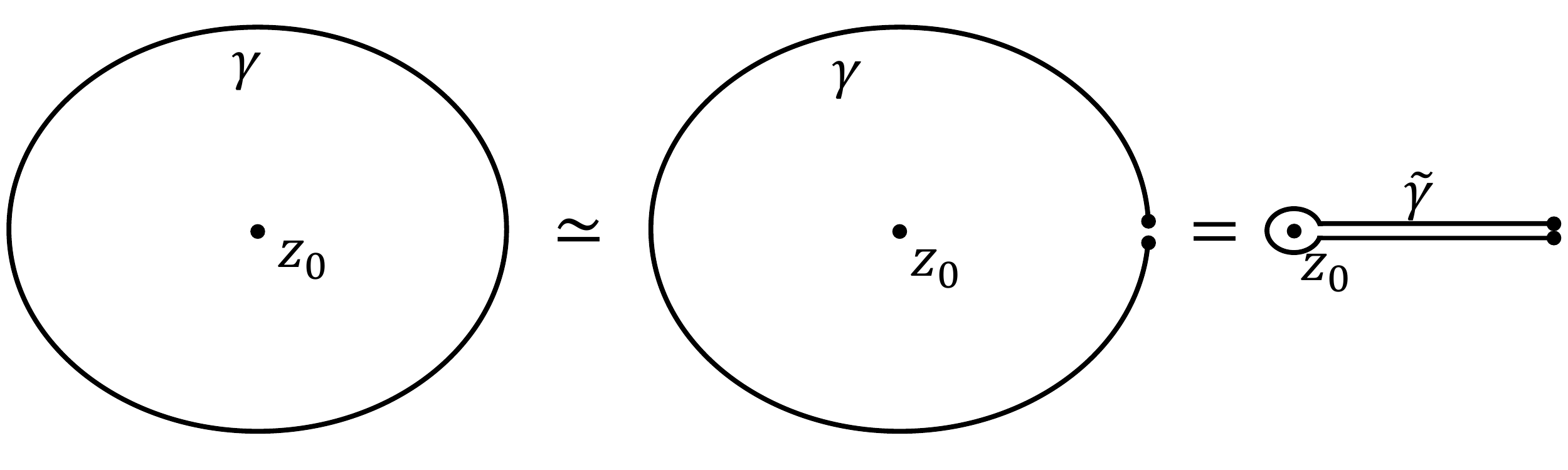}
    \caption{Equivalent contours---the latter two by contour shifting---simplify the integration of \emph{any} closed contour around a singularity (left) to that of a circle arbitrarily close to the singularity (right).}
    \label{fig:figSM1}
\end{figure}
One can use contour-shifting to prove an important integral formula. Consider the closed-contour integral $\oint_\gamma \frac{f(z)}{z - z_0} \,{\rm d}z$, where $f$ is holomorphic on $\gamma$, but there is now a singularity in the integrand. For any arbitrary closed contour $\gamma$, one can follow the prescription of \figref{figSM1}: first make a tiny perforation in the contour, then use that perforation to shift to a modified contour that comprises two straight lines (whose integrals cancel by directionality) and a tiny circle at the origin. On the tiny circle, we can write $f(z) \approx f(z_0)$. On the circle, $z = z_0 + \varepsilon e^{i2\pi t}$, for $t$ from 0 to 1, where $\varepsilon$ is the radius of the circle on $\tilde{\gamma}$, such that
\begin{align}
    \oint_{\tilde{\gamma}} \frac{f(z)}{z - z_0} &\approx f(z_0) \oint_{\tilde{\gamma}} \frac{1}{z - z_0} \,{\rm d}z \nonumber \\
                                                &= f(z_0) \frac{1}{\varepsilon} \oint e^{-i2\pi t}\,{\rm d}\left(\varepsilon e^{i2\pi t}\right) \nonumber \\
                                                &= 2\pi i f(z_0).
                                                \label{eq:CIF}
\end{align}
\Eqref{CIF} is \textbf{Cauchy's integral formula}.

One can take derivatives of \eqref{CIF} with respect to $z_0$ to yield an expression for the first derivative:
\begin{align}
    f'(z_0) = \frac{1}{2\pi i} \oint_\gamma \frac{f(z)}{(z-z_0)^2} \,{\rm d}z,
\end{align}
and more generally \textbf{Cauchy's differentiation formula}:
\begin{align}
    f^{(n-1)}(z_0) = \frac{(n-1)!}{2\pi i} \oint_\gamma \frac{f(z)}{(z-z_0)^n} \,{\rm d}z,
    \label{eq:CDF}
\end{align}

It is then one final step to get from Cauchy's differentiation formula to the residue theorem. Set the integrand in \eqref{CDF} to a function $g(z)$, which has a pole of order $n$ at $z_0$. By a Laurent expansion, can write \emph{any} function with a pole of order $n$ at $z_0$ in this form. Then we have the \textbf{residue theorem}\index{Cauchy residue theorem}:
\begin{align}
    \int_\gamma g(z) \,{\rm d}z = 2\pi i \sum_{\rho} \operatorname{Res}(f;z_0),
\end{align}
where the \textbf{residue} of $f$ at $z_0$ is defined as
\begin{align}
    \operatorname{Res}(f;z_0) = \frac{1}{(n-1)!} \lim_{z\rightarrow z_0} \frac{d^{n-1}}{dz^{n-1}} \left[ (z-z_0)^n f(z) \right].
\end{align}
For $n=1$, a simple pole, the residue is given by
\begin{align}
    \lim_{z\rightarrow z_0} \left[ (z-z_0) f(z) \right].
\end{align}

\bibliographystyle{ieeetr}
\bibliography{reference}

\end{document}